\documentclass[preprints,article,accept,moreauthors,pdftex]{mdpi}
\firstpage{1} 
\pubvolume{1}
\issuenum{1}
\articlenumber{0}
\pubyear{2021}
\copyrightyear{2020}
\datereceived{25 October 2021} 
\dateaccepted{12 December 2021} 
\datepublished{xxx} 
\hreflink{https://doi.org/} 

\usepackage{graphicx}	
\usepackage{amsmath}	
\usepackage{amssymb}	
\usepackage{mathptmx}
\usepackage{mathrsfs}
\usepackage{wasysym}
\usepackage{epsfig}
\usepackage{amsmath}
\usepackage{amssymb}
\usepackage{pdflscape}
\usepackage{pifont}
\usepackage{afterpage}
\usepackage{paralist}
\usepackage{graphicx}
\usepackage{afterpage}
\usepackage{epstopdf}
\usepackage{subfig}
\usepackage{amsfonts}
\usepackage{bm}
\usepackage[flushleft]{threeparttable}
\usepackage{float}
\usepackage{array}
\newcolumntype{P}[1]{>{\centering\arraybackslash}p{#1}}
\usepackage{lscape}
\usepackage{alphalph}

\usepackage[section]{placeins}
\usepackage{url}
\usepackage{natbib}
\usepackage{multicol}
\usepackage{fancyhdr}
\usepackage{enumitem}
\newcommand{\ie}{$i.e.,\;$}
\newcommand{\eg}{$e.g.,\;$}
\newcommand{\viz}{$viz.,\;$}

\Title{Remnant radio galaxy candidates of small angular sizes}
\TitleCitation{Remnant candidates of small angular sizes}
\Author{Veeresh Singh $^{1}$\orcidA{}, Sushant Dutta $^{1,2}$*\orcidB{}, 
Yogesh Wadadekar $^{3}$\orcidC{}, C. H. Ishwara-Chandra $^{3}$\orcidD{}}

\AuthorNames{Veeresh Singh, Sushant Dutta, Yogesh Wadadekar, and C. H. Ishwara-Chandra}

\AuthorCitation{Singh, V.; Dutta, S.; Wadadekar, Y.; Ishwara-Chandra, C.H.}

\address{%
$^{1}$ \quad Physical Research Laboratory, Ahmedabad, 380009, Gujarat, India \\
$^{2}$ \quad Indian Institue of Technology Gandhinagar, Palaj, Gandhinagar, 382355, Gujarat, India \\
$^{3}$ \quad National Centre for Radio Astrophysics - TIFR, 411007, Pune, India \\
}

\corres{Correspondence: veeresh@prl.res.in}
\abstract{Remnant radio galaxies (RRGs), characterized by the cessation of AGN activity, represent 
a short-lived last phase of radio galaxy's life-cycle. Hitherto, searches for RRGs, mainly based on 
the morphological criteria, have identified large angular size sources resulting into a 
bias towards the remnants of powerful FR-II radio galaxies. 
In this study we make the first attempt to perform a systematic search for RRGs of small angular sizes 
($<$30$^{\prime\prime}$) in the XMM-LSS field. 
By using spectral curvature criterion we discover 48 remnant candidates exhibiting strong spectral curvature 
{\ie}${\alpha}_{\rm 150~MHz}^{\rm 325~MHz}$ - ${\alpha}_{\rm 325~MHz}^{\rm 1.4~GHz}$ $\geq$0.5. 
Spectral characteristics at higher frequency regime ($>$1.4 GHz) indicate that some of our remnant candidates 
can depict recurrent AGN activity with an active core. 
We place an upper limit on the remnant fraction ($f_{\rm rem}$) to be 3.9$\%$, which increases 
to 5.4$\%$ if flux cutoff limit of S$_{\rm 150~MHz}$ $\geq$10 mJy is considered. 
Our study unveils, hitherto unexplored, a new population of small-size ($<$200 kpc) remnant candidates 
that are often found to reside in less dense environments and at higher redshifts ($z$) $>$1.0. 
We speculate that a relatively shorter active phase and/or low jet power can be plausible reasons 
for the small size of remnant candidates.
}
\begin{document}
%

\section{Introduction}
\label{sec:Intro}
Radio galaxies, a subclass of Active Galactic Nuclei (AGN), emit copiously at radio wavelengths and exhibit well defined 
radio structures $-$ a radio core, highly collimated outflowing bipolar jets eventually terminating into radio lobes. 
Understanding the evolution of radio galaxies is one of the important aspects of galaxy evolution as 
AGN jet activity influences host galaxy and surrounding inter-galactic-medium (IGM) via feedback processes 
\citep{McNamara07}. Large-area multi-frequency sensitive radio continuum surveys have played a vital role in 
advancing our understanding on the radio galaxies evolution by detecting a large number of sources representing 
different phases of the radio galaxy's life cycle. According to the evolutionary models infancy phase of a radio 
galaxy can be depicted by compact sources with Linear-Angular-Size (LAS) less than a kpc. Based on their 
spectral characteristics these sources are known as High-Frequency-Peakers (HFP) 
and Gigahertz-Peak-Spectrum (GPS) sources, that evolves into 
Compact-Steep-Spectrum (CSS) sources with LAS of a few to tens of kpc \citep{ODeaSaikia21}. 
CSS radio sources can evolve into large-size radio galaxies of a few hundreds of kpc via sustained supply of 
plasma through jets that can remain active for tens of millions of years \citep[see][]{An12,Turner15}. 
Remnant phase begins after the cessation of AGN activity during which jets are no longer sustained and lobes start 
to fade away. In the remnant phase, radio core and jets disappear but the radio lobes can still be detected 
for a time-scale of a few times of 10$^{7}$ years before they disappear due to radiative and adiabatic 
losses \citep{Murgia11}. 
The time-scale over which remnant lobes can be detected is arguably much shorter than the active phase, and hence, 
remnant phase represents a short-lived final phase of radio galaxy evolution \citep{Brienza17}. 
The short-lived remnant phase makes remnant radio galaxies (RRGs) rare objects to be detected. 
\par
In recent times, deep low-frequency radio surveys have been exploited to search for the population of RRGs 
with an expectation to find a large number of remnant sources exhibiting diffuse low-surface-brightness emission of 
steep spectrum.  
However, contrary to the predictions of evolutionary models based on radiative cooling of lobes plasma, the 
fraction of RRGs is found to be as low as 5$\%$ to 10$\%$. 
For instance, using deep 150 LOFAR survey (5$\sigma$ = 0.1 $-$ 2.0 mJy beam$^{-1}$) and deep 6 GHz 
(5$\sigma$ = 0.02 mJy beam$^{-1}$) observations \cite{Mahatma18} identified only 11/127 $<$ 9$\%$ potential 
RRGs with absent-core criterion in the Herschel-ATLAS field. 
In a similar study \cite{Jurlin21} found only 11/158 $<$ 8$\%$ RRGs in the Lockman Hole field.  
Recently, we carried out a search for RRGs in the the XMM-{\em Newton} Large-Scale Structure (XMM-LSS) field using 
deep 325 MHz Giant Metrewave Radio Telescope (GMRT) survey, 150 MHz Low-Frequency Array (LOFAR) survey 
and 1.4 GHz Jansky Very Large Array (JVLA) surveys, and found that the RRGs fraction is even lower 
upto $<$ 5$\%$ if fainter population (S$_{\rm 325~MHz}$ $\geq$ 6.0 mJy) is probed 
(Dutta et al. 2021, {\it submitted}, D21 hereafter). Unlike D21 both \cite{Jurlin21} and \cite{Mahatma18} have 
introduced a high flux density cutoff limits {\ie}S$_{\rm 150~MHz}$ $\geq$ 40 mJy and S$_{\rm 150~MHz}$ $\geq$ 80 mJy 
in their samples, respectively. 
In addition to relatively high flux density cutoff limit, all the previous studies have also introduced a cutoff limit 
of 30$^{\prime\prime}$ $-$ 60$^{\prime\prime}$ on LAS \citep{Saripalli12,Brienza17,Quici21}.
For instance, \cite{Brienza17} identified RRGs using a sample of extended source with 
LAS $\geq$ 60$^{\prime\prime}$, while \cite{Mahatma18} limited their search to the radio sources with 
LAS $\geq$ 40$^{\prime\prime}$ in the 150 MHz LOFAR images.
In the XMM-LSS field, D21 attempted to identify RRGs in a sample of extended sources with 
LAS $\geq$ 30$^{\prime\prime}$ in the 325 MHz GMRT images. 
We note that the requirement of placing a cutoff limit on the LAS arises due the use of morphological 
criteria {\ie}absence of radio core, hotspots, in the radio images of few arcsec resolution. 
Moreover, the cutoff limit placed on the angular size introduces a bias towards large and powerful radio galaxies.  
In their sample of RRGs with LAS $\geq$ 60$^{\prime\prime}$ \cite{Jurlin21} found that all RRGs show 
double-lobe morphology with high radio luminosity (L$_{\rm 150~MHz}$ $>$ 10$^{25}$ W~Hz$^{-1}$) 
suggesting their progenitors to be powerful FR-II radio galaxies. 
\par
We point out that the bias towards the remnants of large powerful radio galaxies continues to exist 
even in case of individual RRGs reported in the literature. 
In fact, individual RRGs were discovered primarily owing to their peculiar radio morphology that appeared 
extended, amorphous, and lacked compact features \citep[{\eg}][]{Tamhane15,Brienza16,Shulevski17,
Duchesne19}. Individual RRGs such as blob1 \citep{Brienza16}, NGC 1534 \citep{Duchesne19}, 
J021659-044920 \citep{Tamhane15}, B2 0924+30 \citep{Cordey87,Jamrozy04,Shulevski17} were identified 
based on their unusual amorphous-shaped large-scale radio morphology of low-surface-brightness emission 
at low-frequencies ($\leq$ 325 MHz). Considering the biases and limitations introduced by the angular size 
cutoff we attempt to search and characterize the nature of 
RRGs of small angular sizes (LAS $<$ 30$^{\prime\prime}$) that have remained unexplored, hitherto.   
We note that small angular size poses difficulty in deciphering radio morphology, and hence, morphological 
criteria cannot be applied to search for small size remnants detected with the radio images of several 
arcsec resolution. Although, remnant characteristics are manifested in the radio spectra in the form of strong 
spectral curvature resulted from the frequency-dependent radiative losses. 
Therefore, we exploit spectral curvature criterion to identify RRGs of small angular sizes. 
In this paper, we use the terms remnants and RRGs interchangeably.   
\\
This paper is structured as follows. In Section~\ref{sec:data} we provide the details of available 
radio and optical data in the XMM-LSS field. 
Section~\ref{sec:identification} describes the selection criteria and the sample of identified remnant candidates. 
In Section~\ref{sec:characteristics} we report the characteristic properties of our remnant candidates and compare 
them with active sources. 
In Section~\ref{sec:smallsize} we attempt to identify plausible reasons for existence of small-size remnants. 
Section~\ref{sec:fraction} is devoted on the discussion for biases that can influence the remnant fraction. 
Section~\ref{sec:conclusions} lists the conclusions of our study. 
\\
In this paper we adopt following cosmological parameters : H$_{\rm 0}$ = 70 km s$^{-1}$ Mpc$^{-1}$, ${\Omega}_{\Lambda}$ = 0.7, 
${\Omega}_{\rm M}$ = 0.3. Radio spectrum is characterized by a power law S$_{\nu}$ $\propto$ ${\nu}^{\alpha}$, where $\alpha$ represents the spectral index. 

\section{Radio and optical surveys in the XMM-LSS field}
\label{sec:data}
\begin{figure}
\includegraphics[angle=0,width=9.0cm,trim={0.3cm 1.7cm 6.0cm 2.0cm},clip]{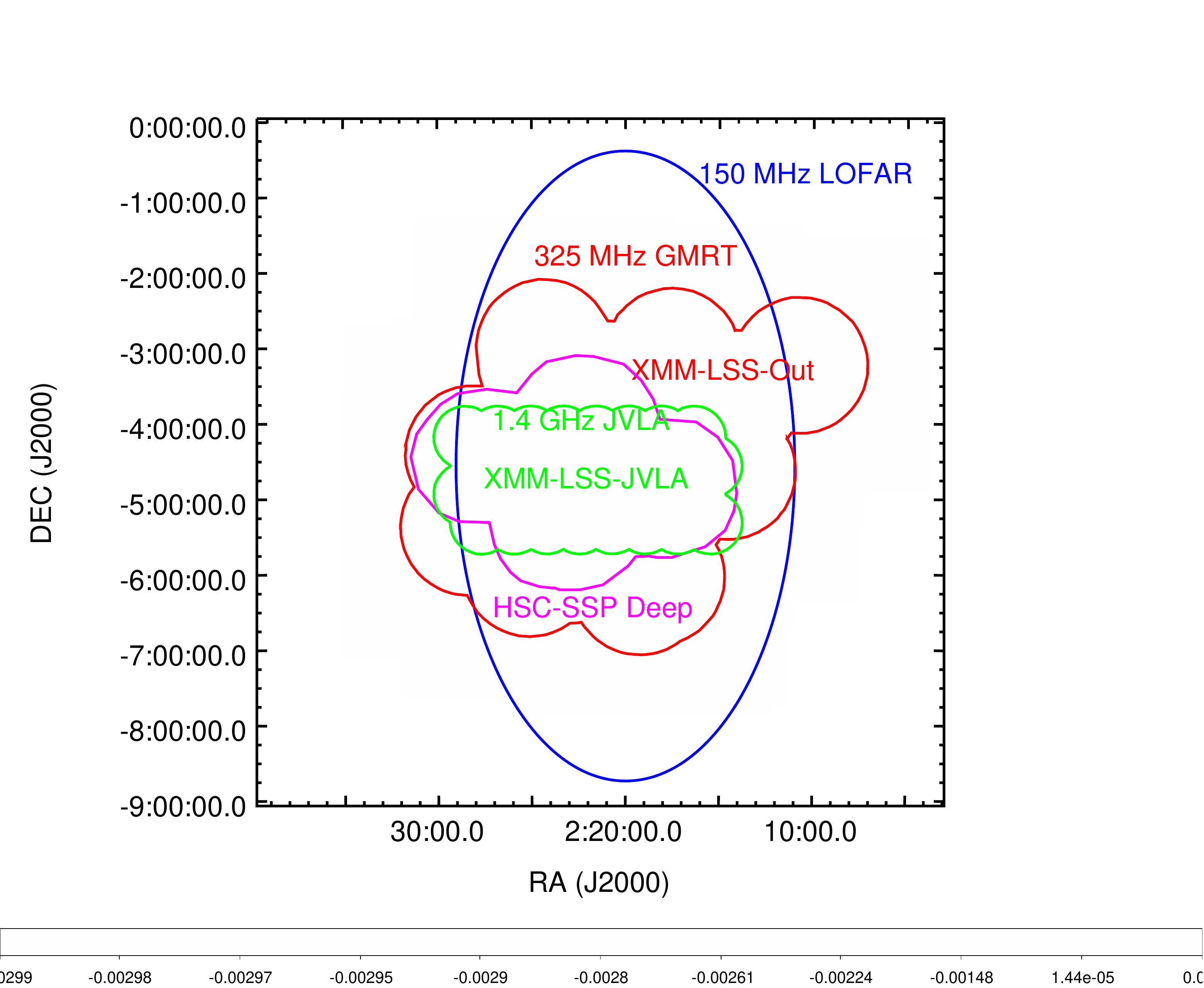}
\caption{Footprints of the 150 MHz LOFAR survey (in Blue), 325 MHz GMRT survey (in Red), 1.4 GHz JVLA (in Green) and the optical HSC-SCP deep component survey (in Magenta) available in the XMM-LSS field. Region covered by the 1.4 JVLA survey is marked as the XMM-LSS-JVLA, while region outside to it survey is marked as the XMM-LSS-Out.}
\label{fig:Footprints} 
\end{figure}

To identify RRG candidates and their host galaxies we use deep multi-frequency radio and optical data available in the XMM-LSS 
field. In our study we mainly utilize 150 MHz LOFAR survey, 325 MHz GMRT survey, 1.4 GHz JVLA survey and 
the Hyper Suprime-Cam Subaru Strategic Program (HSC-SSP) optical survey. 
Figure~\ref{fig:Footprints} shows the footprints of all three radio surveys and of 
the HSC-SSP deep component survey. 
The HSC-SSP wide component survey covers a much larger sky area including the full XMM-LSS field.      
We note that 1.4 GHz JVLA survey overlaps partially with the 325 MHz GMRT survey. 
Therefore, we divide 325 MHz survey region into two parts $-$ (i) 5.0 deg$^{2}$ area covered with the deep 1.4 GHz JVLA survey named 
as the XMM-LSS-JVLA, and (ii) 325 MHz GMRT survey region not covered with the 1.4 GHz JVLA survey named as the XMM-LSS-Out.     
We use relatively less sensitive 1.4 GHz FIRST survey in the XMM-LSS-Out region.
In following subsections we provide brief details of the surveys that are used in our study.      
\subsection{Radio surveys}
\label{sec:radio}
{\it 150 MHz LOFAR survey} : 
150 MHz LOFAR survey covers 27 deg$^{2}$ sky area with elliptical footprints centered at RA = 35$^{\circ}$ DEC = -4$^{\circ}$.5 
in the XMM-LSS field \citep{Hale19}. 
This survey is carried out with the High Band Antenna (HBA; 110-240 MHz) having central frequency at 144 MHz. 
Final image has a median noise-rms of 0.394 mJy beam$^{-1}$ and a noise-rms of 0.28 mJy beam$^{-1}$ in the central region 
and angular resolution of 7$^{\prime\prime}$.5 $\times$ 8$^{\prime\prime}$.5. 
There are a total of 3044 individual radio sources detected in this survey.  
\\
{\it 325 MHz GMRT survey} : 
325 MHz GMRT survey centered at RA = 02h 21m 00s and DEC = -04$^{\circ}$ 30$^{\prime}$ 00$^{\prime\prime}$ 
covers a total sky area of 12.5 deg$^2$ with 16 pointings in the XMM-LSS field.  
This survey was performed in semi-snapshot mode with the legacy GMRT having an instantaneous bandwidth of 32 MHz. 
The final 325 MHz GMRT mosaiced image achieves nearly a uniform sensitivity with an average noise-rms of 
150 $\mu$Jy beam$^{-1}$ and the synthesized beam-size of 10$^{\prime\prime}$.2 $\times$ 7$^{\prime\prime}$.9. 
There are a total of 3739 individual radio sources detected in this survey. We refer reader to \cite{Singh14} for more details on the 325 MHz GMRT survey. 
Assuming a typical spectral index of -0.7, the sensitivity of 325 MHz GMRT survey (5$\sigma$ = 0.75 mJy) 
scales to 1.34 mJy at 150 MHz, that is comparable to the sensitivity of 150 MHz LOFAR survey (5$\sigma$ = 1.4 mJy) in 
the deeper regions. Therefore, combination of 325 MHz GMRT survey and 150 MHz LOFAR survey of similar depths 
and angular resolution will enhance our chances to detect remnants, in particular at the fainter regime. \\
{\it 1.4 GHz JVLA radio survey} : 
1.4 GHz JVLA radio survey carried out with B configuration in the wide-band continuum mode 
with the spectral coverage of 0.994 $-$ 2.018 GHz, 
covers sky area of 5.0 deg$^{2}$ over the near-IR VISTA Deep Extragalactic Observations (VIDEO) survey region 
in the XMM-LSS field \citep{Heywood20}. Final image mosaiced of 32 pointings shows a median noise-rms of 16 $\mu$Jy beam$^{-1}$ and 
an angular resolution of 4$^{\prime\prime}$.5. 
It is the deepest 1.4 GHz survey over the largest area (5.0 deg$^{2}$) in the XMM-LSS field, and it 
detects a total of 5762 individual radio sources above $5\sigma$ flux limit.  
\par
We note that the XMM-LSS field is also surveyed at 610 MHz with GMRT covering a total area of 
30.4 deg$^2$ with a non-uniform noise-rms $-$ 200 $\mu$Jy beam$^{-1}$ in the inner area of 11.9 deg$^{2}$, and 45 $\mu$Jy beam$^{-1}$ 
in the outer area of 18.5 deg$^{2}$ $-$ and it detects 5434 radio sources at $\geq$ 7$\sigma$ level with an angular 
resolution of 6$^{\prime\prime}$.5 \citep{Smolcic18}. In principle, 610 MHz can be used to select sources showing strong spectral curvature 
at frequencies $\geq$ 610 MHz, however, D21 found that 610 MHz flux densities are often underestimated and unreliable. 
Therefore, we prefer not to use 610 MHz GMRT survey. 
Further, shallow 240 MHz GMRT survey (noise-rms of $\sim$2.5 mJy beam$^{-1}$) and 74 MHz VLA survey with a noise-rms 
of 20 $-$ 55 mJy beam$^{-1}$ are also available in the XMM-LSS field. However, majority of our 325 MHz detected sources lack the 
detections in 240 MHz GMRT and 74 MHz VLA surveys owing to the large differences in their sensitivities. Hence, we find that 
74 MHz and 240 MHz surveys are not useful to be combined with deep 150 MHz LOFAR and 325 MHz GMRT survey 
for identifying remnant candidates.   

\subsection{Optical surveys in the XMM-LSS}
\label{sec:optical}
The XMM-LSS field is covered with the HSC-SSP\footnote{https://hsc-release.mtk.nao.ac.jp/doc/}, 
a three-tiered (wide, deep and ultra-deep), multi-band ($g$, $r$, $i$, $z$, $y$ and four narrow-band filters) 
imaging survey. HSC-SSP survey is carried out with the Hyper Suprime-Cam, a wide-field camera, 
installed at the 8.2-m Subaru telescope. 
The wide component of HSC-SSP survey has a sensitivity limit of 26.2$_{\rm -0.3}^{\rm +0.2}$ mag at 5$\sigma$ level 
in $i$ band, while deep component is nearly one magnitude deeper with 5$\sigma$ limit of 
26.9$_{\rm -0.3}^{\rm +0.2}$ mag. 
We note that deep component covers only 7.0 deg$^{2}$ in the XMM-LSS (see Figure~\ref{fig:Footprints}), hence, 
we use HSC-SSP wide component in the regions not covered with the deep component. 
The ultra-deep component is nearly 0.8 magnitude deeper than the deep component and covers only 
1.7 deg$^{2}$ sky area over Subaru XMM-Newton Deep field Survey (SXDS). 
The 5$\sigma$ limiting AB magnitudes are measured within 2$^{\prime\prime}$.0 diameter apertures. 
For our study we use optical data from the HSC-SSP third public data release (PDR3) that provides source catalogues and 
images with the median seeing of 0$^{\prime\prime}$.6. The HSC-SSP PDR3 also provides publicly available 
spectroscopic redshifts (spec-$z$ table in the data access 
website\footnote{https://hsc-release.mtk.nao.ac.jp/doc/index.php/available-data\_pdr3/}). 
We note that spectroscopic redshifts are available only for relative bright sources. 
Therefore, we obtain photometric redshifts from \cite{Schuldt21} who derived photo-$z$ estimates 
of the HSS-SSP objects.  
%
\section{Identification of remnant candidates of small angular sizes}
\label{sec:identification}
In the literature, remnants are identified mostly by using morphological criteria {\viz}absence of radio core and presence of diffuse 
amorphous shaped low-surface-brightness emission \citep{Mahatma18,Quici21,Jurlin21}. For sources of small angular sizes 
(LAS $<$ 30$^{\prime\prime}$) morphological details cannot be deciphered from the images with a typical resolution of 5$^{\prime\prime}$.0 $-$ 10$^{\prime\prime}$.  
Notably, spectral curvature criterion can allow us to identify remnant candidates even among unresolved sources. One of the key examples of such RRGs is J1615+5452 which is identified mainly 
using the spectral curvature criterion and does not reveal much morphological details in the images of 
5$^{\prime\prime}$.0 $-$ 10$^{\prime\prime}$ angular resolution \citep[see][]{Randriamanakoto20}. 
Also, the absence of core supporting the remnant status of J1615+5452 was inferred 
from its non-detection in the moderately deep (5$\sigma$ = 0.45 mJy) 1.4 GHz VLA observations 
of 5$^{\prime\prime}$.0 resolution. We note that the core detection particularly in small sources 
($<$ 30$^{\prime\prime}$) poses requirement for deep (noise-rms less than a few $\mu$Jy), high-resolution 
(of sub-arcsec level or even better) observations at higher frequencies ($>$1.4~GHz). 
Also, we cannot rule out the possibility of existence of a faint 
core falling below the detection limit even in the deep high-frequency observations.  
Hence, we exploit the spectral curvature criterion to identify candidates of remnant sources of small angular sizes. 
\subsection{Spectral curvature selection criterion}
\label{sec:criteria}
Cessation of AGN activity results into the stoppage of jets supplying plasma to the lobes. 
Plasma contained in radio lobes suffers radiative losses without the injection of any fresh plasma. 
Relativistic electrons present in the plasma lose their energy via synchrotron emission as well as 
Inverse Compton (IC) scattering with the Cosmic Microwave Background (CMB) photons \citep{Komissarov94}. 
Since electrons of higher-energy lose their energy faster than the low-energy electrons, a spectral break 
in the power law radio spectrum develops such that the spectrum becomes steeper at higher frequency 
above the break frequency (${\nu}_{\rm b}$), while spectrum continues to exhibit original spectral index 
below ${\nu}_{\rm b}$.
Thus, radio spectrum of a remnant source can be represented by a broken power law or a curved power law 
that exhibits spectral index same to that of injected plasma ${\alpha}_{\rm inj}$, typically in the range -0.5 to -1.0, below ${\nu}_{\rm b}$, 
and a steeper spectral index ${\alpha}_{\rm inj}$ - 0.5 above ${\nu}_{\rm b}$ \citep{Blandford78,Murgia11}.
According to spectral aging models, with time, ${\nu}_{\rm b}$ progressively shifts towards lower frequencies \citep{Jaffe73}. 
To identify radio sources showing curved radio spectrum we examine the difference between low-frequency 
and high-frequency spectral indices {\ie}${\Delta}{\alpha}$ = ${\alpha}_{\rm low}$ - ${\alpha}_{\rm high}$. 
In fact, ${\Delta}{\alpha}$ depicts spectral curvature parameter (SCP) $-$ an indicator of evolutionary stage 
of a radio galaxy, and the typical value of ${\Delta}{\alpha}$ is $\geq$ 0.5 for a RRG \citep[see][]{Murgia11}. 
In our study, we define ${\Delta}{\alpha}$ = ${\alpha}_{\rm 150~MHz}^{\rm 325~MHz}$ - ${\alpha}_{\rm 325~MHz}^{\rm 1.4~GHz}$, 
and consider ${\Delta}{\alpha}$ $\geq$ 0.5 as a characteristic signature of remnant.   
We note that the spectral curvature criterion selects only a fraction of remnants that show 
strong curvature ($\geq$ 0.5) within the considered frequency coverage of 150 MHz $-$ 1.4 GHz. 
Remnants with spectral break (${\nu}_{\rm b}$) 
falling outside the frequency window of 150~MHz $-$ 1.4~GHz and even below 325 MHz would be missed in our study.
In following subsection we describe the identification of remnant candidates. 
\subsection{325 MHz GMRT sources of small angular sizes}
\label{sec:sample}
Our initial sample is consisted of 2513 radio sources detected in the 325 MHz GMRT survey 
with the signal-to-noise ratio (SNR) $\geq$ 7 and LAS $<$ 30$^{\prime\prime}$. 
The SNR cutoff is chosen to avoid any contamination from spurious sources. 
While, cutoff on the LAS is applied owing to the fact that the sample of extended radio sources 
with LAS $\geq$ 30$^{\prime\prime}$ is already 
probed to search for RRGs reported in D21. 
To use spectral curvature criterion we obtain low-frequency (${\alpha}_{\rm 150~MHz}^{\rm 325~MHz}$) 
and high-frequency (${\alpha}_{\rm 325~MHz}^{\rm 1.4~GHz}$) spectral indices of 325 MHz GMRT sources by 
finding their counterparts at 150 MHz and 1.4 GHz. Table~\ref{tab:steps} lists the number of 325 MHz sources 
with counterparts at 150 MHz and 1.4 GHz.
\subsubsection{Counterparts at 150 MHz and 1.4 GHz}
We search 150 MHz counterparts of 325 MHz GMRT sources using 150 MHz LOFAR survey, whenever available, 
otherwise a relatively shallow 150 MHz TIFR GMRT Sky Survey (TGSS; \citep{Intema17}) is used. 
We find that 2179 out of 2513 (86.7$\%$) sources fall within the LOFAR survey region owing to a substantially large overlap between 
325 MHz GMRT survey and 150 MHz LOFAR survey (see~Figure~\ref{fig:Footprints}). 
Cross-matching of 325 MHz GMRT sources to 150 MHz source catalogue with a search radius of 15$^{\prime\prime}$ gives only 1480 sources. 
We note that a larger search radius of 15$^{\prime\prime}$ is considered due to large positional uncertainties 
(a few arcsec) associated with faint diffuse sources detected at both frequencies. 
In fact, majority of sources are found to be matched within 5$^{\prime\prime}$ radius. 
The number of cross-matched sources by chance is only 1.6$\%$, if source density of 300 deg$^{-2}$ in the 325 GMRT survey 
is considered. The non-detection of 699/2179 (32$\%$) of 325 MHz GMRT sources in the LOFAR survey 
can be understood due to 
their low flux densities {\ie}fainter source population. We find that the 325 MHz flux density distribution for the  
non-detected sources peaks around 1.0 mJy and most of the sources are fainter than 2.0 mJy. 
\begin{table}
\caption{Selection steps for remnant candidates}
\label{tab:steps}
\centering
\begin{tabular}{lcc} \hline 
 Sample           &    Criteria                           &    Size       \\ \hline
Detected at 325 MHz GMRT & SNR $\geq$ 7 and size $<$ 30$^{\prime\prime}$ & 2513     \\
Detected at 150 MHz      &                                               & 1516      \\
LOFAR                    &                                               & 1480/2179 \\
TGSS (outside LOFAR region) &                                            & 36/334    \\
Detected at 325 MHz and 150 MHz  & ${\alpha}_{\rm 150~MHz}^{\rm 325 MHz}$ constrained & 1516    \\
Detected at 1.4 GHz &                                                  &  1160       \\     
NVSS         &                                                  & 599/1516  \\
JVLA         &                                                  & 430/434  \\
FIRST (outside JVLA region) &                            & 131/483  \\
150 MHz, 325 MHz and 1.4 GHz & ${\alpha}_{\rm 150~MHz}^{\rm 325 MHz}$  and ${\alpha}_{\rm 325~MHz}^{\rm 1.4~GHz}$ &  1160 \\
150 MHz, 325 MHz but no 1.4 GHz &   ${\alpha}_{\rm 150~MHz}^{\rm 325 MHz}$ but ${\alpha}_{\rm 325~MHz}^{\rm 1.4~GHz}$ upper limits & 356    \\                  
Remnant candidates & ${\Delta}{\alpha}$ $\geq$ 0.5 and ${\alpha}_{\rm 150~MHz}^{\rm 325~MHz}$ $\leq$ -0.5   & 48/1516 (3.2$\%$)  \\ \hline
Redshift estimates & spec-$z$                                   & 353   \\
                   & photo-$z$                                  & 505   \\
                   & no redshift                                & 658   \\  \hline
\end{tabular}
\end{table} 
For 334 radio sources falling outside the LOFAR survey region we searched 150 MHz counterparts using TGSS source catalogue having flux density limit of 24.5 mJy at 7$\sigma$. We find that only 36 relatively bright 
sources (S$_{\rm 325~MHz}$ $\geq$ 11 mJy) show counterparts in the TGSS. 
Thus, we have a total of 1516 sources detected at both 325 MHz and 150 MHz frequencies, providing the estimates of 
spectral index between 150 MHz $-$ 325 MHz (${\alpha}_{\rm 150~MHz}^{\rm 325~MHz}$). 
We do not consider sources for which low-frequency spectral indices are unconstrained {\ie}no detected counterparts 
at 150~MHz. Hence, our study is limited to the sample of 1516 radio sources detected at the both 325 MHz and 150 MHz frequencies.
\par
To estimate high-frequency (325 MHz $-$ 1.4 GHz) spectral index we attempt to find 1.4 GHz counterparts for 1516 sources by 
using the NVSS, whenever available, otherwise JVLA and FIRST surveys are used. 
We prefer to use the NVSS survey due to its large beam-size (45$^{\prime\prime}$) that is more effective in 
capturing low-surface-brightness-emission often associated with remnant sources. 
We find that only 599/1516 sources show 1.4 GHz counterparts in the NVSS catalogue within a search radius 
of 15$^{\prime\prime}$. Low detection rate in the NVSS is due to its higher detection limit of 
2.5 mJy at 5$\sigma$, that corresponds 
to nearly 7.0 mJy at 325 MHz, if a typical spectral index of -0.7 is assumed. 
We caution that the large beam-size of NVSS (45$^{\prime\prime}$) can 
occasionally suffer contamination from the neighboring sources which can result spectral index flatter than the true value. 
For 917 sources with no counterparts in the NVSS we search for the 1.4 GHz counterparts using JVLA survey, if available, 
otherwise the FIRST survey.  
We find that only 434 sources fall within the JVLA survey region, and 430/434 sources show counterparts in the 1.4 GHz JVLA survey. 
High detection rate in the JVLA survey is due to its high sensitivity (0.08 mJy at 5$\sigma$). 
Among remaining 483 sources we find 1.4 GHz counterparts of only 131 sources in the FIRST survey. 
We note that, for 356 sources with no detected counterparts at 1.4 GHz, we place an upper limit of 2.5 mJy based the NVSS detection limit, 
if source is resolved, otherwise a more stringent upper limit of 1.0 mJy is placed based the FIRST survey.
\subsubsection{Remnant candidates identified from ${\alpha}_{\rm low}$ versus ${\alpha}_{\rm high}$ diagnostic plot}
We use ${\alpha}_{\rm low}$ versus ${\alpha}_{\rm high}$ diagnostic plot to identify remnant candidates exhibiting 
strong spectral curvature. With our spectral coverage limited to only three frequencies {\ie}150 MHz, 325 MHz and 1.4 GHz, we 
consider ${\alpha}_{\rm low}$ = ${\alpha}_{\rm 150~MHz}^{\rm 325~MHz}$ and ${\alpha}_{\rm high}$ = ${\alpha}_{\rm 325~MHz}^{\rm 1.4~GHz}$. 
\begin{figure*} 
\includegraphics[angle=0,width=8.0cm,trim={0.0cm 0.0cm 0.0cm 0.0cm},clip]{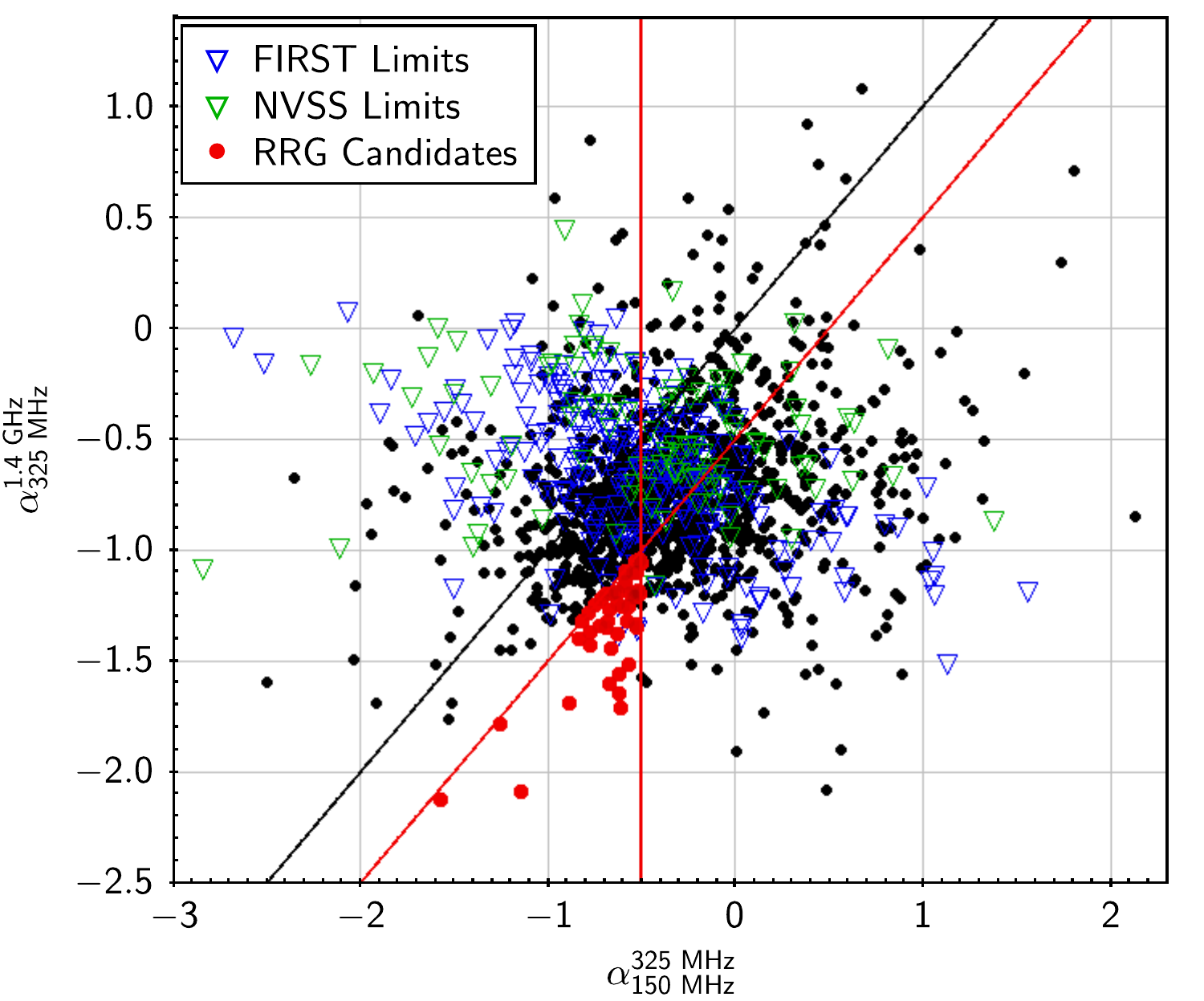}
{\includegraphics[angle=0,width=8.0cm,trim={0.0cm 0.0cm 0.0cm 0.0cm},clip]{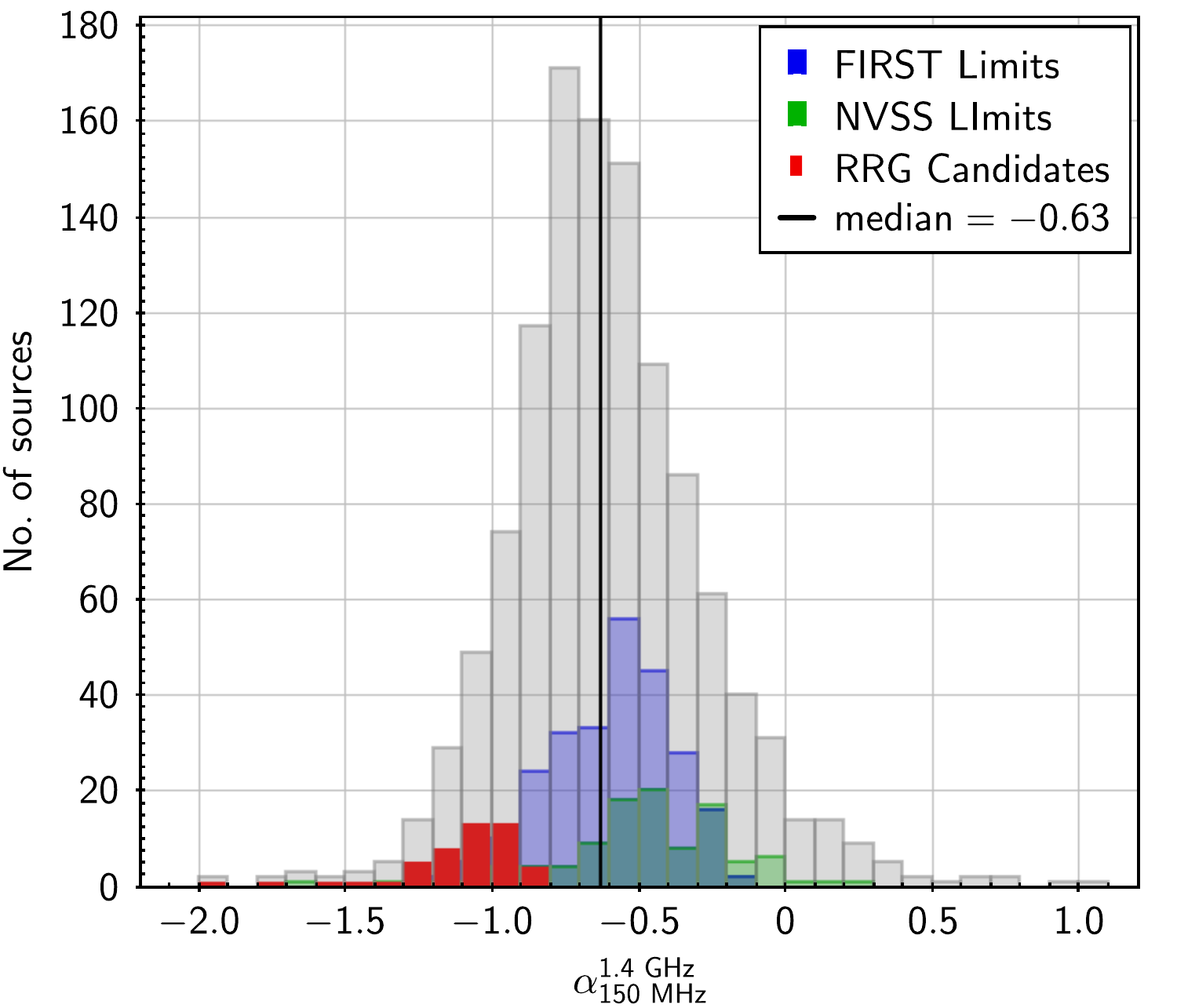}}
\caption{{\it Left panel}: Diagnostic plot of ${\alpha}_{\rm 150~MHz}^{\rm 325~MHz}$ versus ${\alpha}_{\rm 325~MHz}^{\rm 1.4~GHz}$. 
Vertical red line represents ${\alpha}_{\rm 150~MHz}^{\rm 325~MHz}$ = -0.5, while diagonal black and red lines depict 
${\alpha}_{\rm 150~MHz}^{\rm 325~MHz}$ = ${\alpha}_{\rm 325~MHz}^{\rm 1.4~GHz}$ and 
${\alpha}_{\rm 150~MHz}^{\rm 325~MHz}$ = ${\alpha}_{\rm 325~MHz}^{\rm 1.4~GHz}$ + 0.5, respectively. 
{\it Right panel}: Histogram of spectral index between 150 MHz and 1.4 GHz.}
\label{fig:AlphaLowVsAlphaHigh} 
\end{figure*}
%
Figure~\ref{fig:AlphaLowVsAlphaHigh} ({\it Left panel}) shows 
${\alpha}_{\rm 150~MHz}^{\rm 325~MHz}$ versus ${\alpha}_{\rm 325~MHz}^{\rm 1.4~GHz}$ plot for our 
sample of 1516 sources that include upper limits on ${\alpha}_{\rm 325~MHz}^{\rm 1.4~GHz}$ for 356 sources. 
We find that our sources exhibit concentration at ${\alpha}_{\rm 150~MHz}^{\rm 325~MHz}$ = -0.36, 
and ${\alpha}_{\rm 325~MHz}^{\rm 1.4~GHz}$ = -0.79 that are the median values of respective spectral index distributions. 
The substantial difference between the median spectral indices at lower and higher frequencies suggests spectral aging 
even in active sources. This fact is also evident from a systematic shift for a majority of sources from the line representing 
${\alpha}_{\rm 150~MHz}^{\rm 325~MHz}$ = ${\alpha}_{\rm 325~MHz}^{\rm 1.4~GHz}$ (black diagonal line) towards 
${\alpha}_{\rm 150~MHz}^{\rm 325~MHz}$ = ${\alpha}_{\rm 325~MHz}^{\rm 1.4~GHz}$ + 0.5 (red diagonal line) 
in Figure~\ref{fig:AlphaLowVsAlphaHigh} ({\it Left panel}). 
Sources showing strong spectral curvature of ${\Delta}{\alpha}$ 
(= ${\alpha}_{\rm 150~MHz}^{\rm 325~MHz}$ - ${\alpha}_{\rm 325~MHz}^{\rm 1.4~GHz}$) $\geq$ 0.5 fall beyond the  
${\alpha}_{\rm 150~MHz}^{\rm 325~MHz}$ = ${\alpha}_{\rm 325~MHz}^{\rm 1.4~GHz}$ + 0.5 line. 
We note that while identifying remnant candidates of strong spectral curvature we 
avoid radio sources showing flat and inverted spectral index at lower frequency {\ie}${\alpha}_{\rm 150~MHz}^{\rm 325~MHz}$ $>$ -0.5, 
that is understood to arise from Self-Synchrotron Absorption (SSA) caused by an active core. 
Thus, we identify sources showing spectral curvature (${\Delta}{\alpha}$) $\geq$ 0.5, 
and low-frequency spectral index steeper than -0.5 (sources falling within the region bounded by 
${\alpha}_{\rm 150~MHz}^{\rm 325~MHz}$ = ${\alpha}_{\rm 325~MHz}^{\rm 1.4~GHz}$ + 0.5 line and 
${\alpha}_{\rm 150~MHz}^{\rm 325~MHz}$ $=$ -0.5 line in Figure~\ref{fig:AlphaLowVsAlphaHigh}, ({\it Left panel})) 
as the potential candidates of remnants. 
We find only 48 remnant candidates with 06 sources having only upper limit on the 1.4 GHz flux density.
Table~\ref{tab:sample} shows the list of our 48 remnant candidates. 
We caution that strong spectral curvature seen for our remnant candidates can be affected 
by the resolution bias {\ie}unlike 
150 MHz and 325 MHz, 1.4 GHz observations especially from the JVLA and FIRST can underestimate the flux density 
by missing the detection of diffuse low-surface-brightness emission. 
\section{Characteristics of RRG candidates}
\label{sec:characteristics}
In this section we describe characteristics properties of our remnant candidates and compare them with active sources.  
\subsection{Broad-band spectral index (${\alpha}_{\rm 150~MHz}^{\rm 1.4~GHz}$)}
\label{sec:index}
Active sources are generally expected to show power law radio spectrum with a spectral index in the range of -0.5 to -1.0. 
Spectral turn over caused by SSA is often seen in active sources with 
a peak (${\nu}_{\rm p}$) $\leq$ 1.0 GHz, that shifts towards lower frequencies 
as a radio source evolves and becomes lobe-dominated \cite{An12}.       
In contrast, remnant sources exhibit power law spectrum with a spectral break that progressively shifts towards 
lower frequency as the remnant source evolves with time \citep{Murgia11}.  
Therefore, spectral index measured over a broad range can give us a clue about the evolutionary stage of a radio source.
We estimate spectral index between 150 MHz and 1.4 GHz (${\alpha}_{\rm 150~MHz}^{\rm 1.4~GHz}$), 
the widest spectral coverage available, for our sample sources.   
Figure~\ref{fig:AlphaLowVsAlphaHigh} ({\it Right panel}) shows the spectral index (${\alpha}_{\rm 150~MHz}^{\rm 1.4~GHz}$) 
distributions of our remnant candidates and active sources. 
Spectral index (${\alpha}_{\rm 150~MHz}^{\rm 1.4~GHz}$) for our remnant candidates is found to be distributed in the range 
of -1.94 to -0.86 with a median value of -1.04. While, active sources have ${\alpha}_{\rm 150~MHz}^{\rm 1.4~GHz}$ is 
the range of -1.91 to 1.09 with a median value of -0.61. 
Thus, it is evident that our remnant candidates show systematically steeper spectral index than that for the active sources. 
The two-sample KS test shows that the distributions of ${\alpha}_{\rm 150~MHz}^{\rm 1.4~GHz}$ for our remnant candidates 
and active sources are different (see Table~\ref{tab:stat}). 
We caution that the systematically steeper spectral index of our remnant candidates can be the result of selection criteria bias 
{\ie}strong spectral curvature (${\Delta}{\alpha}$ $\geq$ 0.5) and ${\alpha}_{\rm 150~MHz}^{\rm 325~MHz}$ $<$ -0.5. 
In Figure~\ref{fig:Spectra} we show three examples of the radio spectra of our remnant candidates. 
We find that the radio spectra of our remnant candidates are better fitted by a curved power law that is defined as 
S$_{\nu}$ = S$_{\rm 0}$ ${({\nu}/{\nu}_{\rm 0})}^{\alpha}$ e$^{q(ln{\nu})^2}$, 
where $q$ parameterises the curvature of spectrum such that $q< 0$ represents a convex spectrum \citep[{see}][]{Quici21}.
For optically-thin synchrotron emission arising from radio lobes, typical value of $q$ ranges over $-0.2 \leq q \leq 0$. 
We find that values of $q$ parameter of our remnant candidates is generally $<$ -0.2 consistent with an optically-thin 
emission from the relic lobes.
\par
We point out that the two point spectral index ${\alpha}_{\rm 150~MHz}^{\rm 325~MHz}$ of our remnant candidates is similar to 
that found for the large-size remnants. For instance, large-size remnant candidates reported in D21 
show ${\alpha}_{\rm 150~MHz}^{\rm 1.4~GHz}$ distributed in the range of 
-1.71 to -0.74 with a median value of -1.02. It is worth to note that the majority (18/24) of remnant candidates in D21 are identified 
using mainly morphological criteria. Also, using only morphological criterion, \citep{Mahatma18} found that the 
spectral index (${\alpha}_{\rm 150~MHz}^{\rm 1.4~GHz}$ ) of their remnant candidates is distributed in the range of -1.5 to -0.5 
with a median value of -0.97. Thus, we find that the spectral index distribution for our small-size remnant candidates 
is similar to that for the large-size remnant candidates despite being selected using different criteria. 
The similar spectral index distributions for our remnant candidates and large-size remnants selected from morphological criteria 
strengthens the remnant status of our candidates. 
%
%
\begin{table} 
\caption{Sample of remnant candidates}
\label{tab:sample}
\hskip-5.0cm
\scalebox{0.65}{
\begin{tabular}{lccccccccccccc} 
\hline 
Source & S$_{\rm 150~MHz}$  & S$_{\rm 325~MHz}$ & S$_{\rm 1.4~GHz}$ & S$_{\rm 3.0~GHz}$ & ${\alpha}_{\rm 150~MHz}^{\rm 325~MHz}$ & ${\alpha}_{\rm 325~MHz}^{\rm 1.4~GHz}$ & ${\Delta}{\alpha}$ & ${\alpha}_{\rm 150~MHz}^{\rm 1.4~GHz}$  & ${\alpha}_{\rm 1.4~GHz}^{\rm 3.0~GHz}$ & LAS  & $z$ & logL$_{\rm 150~MHz}$ \\
Name    & (mJy)         & (mJy)         &  (mJy)    &   (mJy)  &      &      &     &    &     & $^{\prime\prime}$ (kpc) &   & (W~Hz$^{-1}$)   \\ \hline
GMRT020750-030455 & 189.6$\pm$19.5& 106.4$\pm$0.6 & 17.1$\pm$0.7 & 4.39$\pm$0.57 & -0.75$\pm$0.13 & -1.25$\pm$0.03 & 0.51 & -1.08$\pm$0.05 &  -1.78$\pm$0.18 & 24.9 (201.2) & 1.041$\pm$0.105    &  27.06 \\
GMRT021109-033244 & 31.0$\pm$5.5  & 17.1$\pm$0.2  & 2.29$\pm$0.4 & 2.03$\pm$0.33 & -0.77$\pm$0.23 & -1.37$\pm$0.12 & 0.60 & -1.16$\pm$0.11 & -0.16$\pm$0.31  & $<$10 ($<$80.1) & $>$1.0 & $>$26.26 \\
GMRT021146-041402 & 34.3$\pm$0.8  & 20.6$\pm$0.35 & 2.5$\pm$0.5  &         & -0.66$\pm$0.04 & -1.44$\pm$0.14 & 0.78 & -1.17$\pm$0.09 &       & 21.1 (168.9) & $>$1.0  & $>$26.31 \\
GMRT021408-053456 & 10.6$\pm$0.5  & 7.0$\pm$0.6   & $<$1.0  &          & -0.53$\pm$0.13 & $<$-1.33       & $>$0.80 & $<$-1.06  &        & $<$10 ($<$57) & 0.442$\pm$0.036  & 24.89 \\
GMRT021446-053941 & 29.5$\pm$0.8  & 19.6$\pm$0.56 & 3.9$\pm$0.4 &         & -0.53$\pm$0.05 & -1.11$\pm$0.07 & 0.58 & -0.91$\pm$0.05 &        & $<$10 ($<$83.8) & 1.30$\pm$0.102   &  26.43  \\
GMRT021504-030420 & 31.0$\pm$0.6  & 16.3$\pm$0.18 & 2.1$\pm$0.4 & 1.50$\pm$0.29 & -0.83$\pm$0.03 & -1.40$\pm$0.13 & 0.57 & -1.20$\pm$0.09 & -0.44$\pm$0.35 & $<$10 ($<$76.1) & 0.834$\pm$0.065   &  26.07  \\
GMRT021528-044045 & 29.9$\pm$1.0  & 20.3$\pm$0.30 & 4.4$\pm$0.02 &         & -0.51$\pm$0.05 & -1.05$\pm$0.01 & 0.54 & -0.86$\pm$0.01 &       & 24 (119) & 0.353$\pm$0.023  &  25.08 \\
GMRT021536-045220 & 3.1$\pm$0.5   & 2.01$\pm$0.18 & 0.34$\pm$0.01 &         & -0.55$\pm$0.25 & -1.21$\pm$0.06 & 0.67 & -0.98$\pm$0.08 &       & $<$10 ($<$80.1) & $>$1.0 & $>$25.19 \\
GMRT021548-035934 & 20.8$\pm$0.9  & 8.61$\pm$0.45 & 0.41$\pm$0.01 &         & -1.14$\pm$0.09 & -2.09$\pm$0.04 & 0.95 & -1.76$\pm$0.02 &       & 17.7 (124.8) & 0.677$\pm$0.042  & 25.79 \\
GMRT021555-041245 & 4.1$\pm$0.5   & 2.56$\pm$0.26 & 0.26$\pm$0.01 &         & -0.62$\pm$0.19 & -1.56$\pm$0.06 & 0.95 & -1.24$\pm$0.05 &       & $<$10 ($<$84.4) & 1.413$\pm$0.132  &  25.79 \\
GMRT021603-025647 & 20.8$\pm$0.6  & 11.1$\pm$0.20 & 1.60$\pm$0.15 &        & -0.81$\pm$0.04 & -1.32$\pm$0.07 & 0.51 & -1.15$\pm$0.04 &       & $<$10 ($<$83.1) & 1.221$\pm$0.154   &   26.30 \\
GMRT021646-051004 & 2.9$\pm$0.5   & 1.93$\pm$0.12 & 0.38$\pm$0.01 &        & -0.54$\pm$0.23 & -1.10$\pm$0.04 & 0.57 & -0.91$\pm$0.07 &       & $<$10 ($<$80.1) &  $>$1.0 & $>$25.16 \\
GMRT021702-060327 & 350.3$\pm$4.0 & 132.8$\pm$0.4 & 9.8$\pm$0.5 & 4.73$\pm$0.54 & -1.25$\pm$0.02 & -1.78$\pm$0.03 & 0.53 & -1.60$\pm$0.02 & -0.96$\pm$0.16 & 25.4 (203.4) & $>$1.0  &  $>$27.44 \\
GMRT021706-031513 & 36.7$\pm$0.7  & 23.3$\pm$0.43 & 4.70$\pm$0.5 &     & -0.58$\pm$0.04 & -1.09$\pm$0.07 & 0.51 & -0.92$\pm$0.05 &      & $<$10 ($<$80.1) & $>$1.0  & $>$26.26 \\
GMRT021759-061642 & 19.5$\pm$0.4  & 13.2$\pm$0.62 & 2.8$\pm$0.6 &         & -0.50$\pm$0.07 & -1.06$\pm$0.15 & 0.56 & -0.87$\pm$0.10 &       & 29 (232.3) & $>$1.0 &  $>$25.97 \\
GMRT021836-035711 & 25.9$\pm$0.5  & 17.41$\pm$0.13& 3.10$\pm$0.5 & 2.46$\pm$0.36 & -0.52$\pm$0.03 & -1.18$\pm$0.11 & 0.67 & -0.95$\pm$0.07 & -0.30$\pm$0.29 & $<$10.0 ($<$84.1) & 1.362$\pm$0.145  & 26.44       \\
GMRT021904-063436 & 48.1$\pm$0.8  & 28.3$\pm$0.3  & 3.94$\pm$0.16 &     & -0.69$\pm$0.02 & -1.35$\pm$0.03 & 0.66 & -1.12$\pm$0.02 &     & 17.3 (132.6) & 0.853$\pm$0.0652 & 26.26 \\
GMRT021907-061611 & 36.5$\pm$0.8  & 21.69$\pm$0.20& 3.40$\pm$0.5 & 2.26$\pm$0.25 & -0.67$\pm$0.03 & -1.27$\pm$0.10 & 0.59 & -1.06$\pm$0.07 & -0.53$\pm$0.24 & $<$10 ($<$80.1) & $>$1.0 & $>$26.29 \\
GMRT021917-042654 & 9.7$\pm$0.5   & 2.87$\pm$0.16 & 0.12$\pm$0.01 &        & -1.58$\pm$0.09 & -2.13$\pm$0.09 & 0.55 & -1.94$\pm$0.03 &       & $<$10 ($<$80.1) & $>$1.0 & $>$25.99 \\
GMRT021926-051535 & 20.4$\pm$0.7  & 13.3$\pm$0.37 & 2.3$\pm$0.5 &         & -0.55$\pm$0.06 & -1.20$\pm$0.15 & 0.64 & -0.98$\pm$0.10 &       & 20.1 (160.9) & $>$1.0 & $>$26.02 \\
GMRT022024-040240 & 6.4$\pm$0.5   & 3.9$\pm$0.18  & 0.68$\pm$0.01 &        & -0.63$\pm$0.11 & -1.20$\pm$0.03 & 0.56 & -0.99$\pm$0.03 &       & $<$10 ($<$84.2) & 1.377$\pm$0.136 &  25.87 \\
GMRT022106-043925 & 3.6$\pm$0.4   & 2.3$\pm$0.16  & 0.40$\pm$0.01  &        & -0.54$\pm$0.17 & -1.21$\pm$0.05 & 0.68 & -0.98$\pm$0.05 &       & $<$10 ($<$79.9) & 0.993$\pm$0.0862  &  25.25 \\
GMRT022127-063808 & 29.4$\pm$0.6  & 18.6$\pm$0.35 & 3.40$\pm$0.4 & 1.95$\pm$0.24 & -0.59$\pm$0.04 & -1.16$\pm$0.08 & 0.57 & -0.96$\pm$0.05 & -0.73$\pm$0.22 & $<$10 ($<$84.5) & 1.446$\pm$0.144 & 26.57  \\
GMRT022145-032930 & 34.8$\pm$0.6  & 19.0$\pm$0.19 & 2.90$\pm$0.5 & 2.68$\pm$0.53 & -0.78$\pm$0.02 & -1.29$\pm$0.12 & 0.51 & -1.11$\pm$0.08 & -0.10$\pm$0.34 & 18 (144.2) & $>$1.0  &  $>$26.29 \\
GMRT022152-053619 & 9.7$\pm$0.5   & 6.2$\pm$0.20  & 0.67$\pm$0.01 &        & -0.57$\pm$0.08 & -1.52$\pm$0.03 & 0.95 & -1.19$\pm$0.02 &      & $<$10 ($<$76.4) & 0.843$\pm$0.0765  &  25.57 \\
GMRT022207-040055 & 4.1$\pm$0.5   & 2.75$\pm$0.21 & 0.48$\pm$0.01 &        & -0.51$\pm$0.18 & -1.20$\pm$0.05 & 0.69 & -0.96$\pm$0.05 &       & $<$10 ($<$80.1) & $>$1.0 &  $>$25.32 \\
GMRT022211-054906 & 182.9$\pm$2.1 & 121.4$\pm$0.26& 21.4$\pm$0.8 & 8.31$\pm$0.23 & -0.53$\pm$0.01 & -1.19$\pm$0.03 & 0.66 & -0.96$\pm$0.02 & -1.24$\pm$0.06 & $<$10 ($<$80.1) & $>$1.0 & $>$26.97 \\
GMRT022227-040719 & 18.4$\pm$0.5  & 10.6$\pm$0.24 & 1.79$\pm$0.01 &        & -0.71$\pm$0.05 & -1.22$\pm$0.02 & 0.50 & -1.04$\pm$0.01 &       & 16.6 (135.4) & 1.087$\pm$0.104 & 26.09 \\
GMRT022250-031152 & 39.2$\pm$1.0  & 23.5$\pm$0.36 & 4.10$\pm$0.50 &        & -0.66$\pm$0.04 & -1.19$\pm$0.08 & 0.54 & -1.01$\pm$0.06 &       & $<$10 ($<$80.1) & $>$1.0 & $>$26.31 \\
GMRT022301-060627 & 137.3$\pm$1.5 & 88.6$\pm$0.35 & 14.2$\pm$0.60 & 4.88$\pm$0.32 & -0.57$\pm$0.01 & -1.25$\pm$0.03 & 0.68 & -1.01$\pm$0.02 & -1.40$\pm$0.10 & 27.3 (218.6) & $>$1.0 & $>$26.86 \\
GMRT022302-024656 & 10.8$\pm$1.1  & 7.2$\pm$0.35  & $<$1.0 &           & -0.53$\pm$0.15 &$<$-1.35        &$>$0.83 &$<$-1.07  &        & $<$10 ($<$76.1) & 3.117$\pm$0.318  & 27.01 \\
GMRT022302-042850 & 12.3$\pm$0.6  & 7.2$\pm$0.13  & 1.26$\pm$0.01 &        & -0.69$\pm$0.06 & -1.19$\pm$0.01 & 0.50 & -1.02$\pm$0.02 &       & $<$10 ($<$82.7) & 1.176$\pm$0.112 &  25.99 \\
GMRT022338-045418 & 7.04$\pm$0.3  & 4.42$\pm$0.47 & 0.70$\pm$0.01 &        & -0.60$\pm$0.15 & -1.25$\pm$0.07 & 0.65 & -1.03$\pm$0.02 &       & 15.3 (122.5) & $>$1.0 & $>$25.57  \\
GMRT022413-044643 & 7.97$\pm$0.7  & 4.02$\pm$0.15 & 0.34$\pm$0.01 &        & -0.89$\pm$0.14 & -1.69$\pm$0.03 & 0.80 & -1.41$\pm$0.04 &       & $<$10 ($<$80.1) & $>$1.0  & $>$25.74 \\
GMRT022433-043709 & 4.1$\pm$0.5   & 2.47$\pm$0.11 & 0.43$\pm$0.01 &        & -0.65$\pm$0.17 & -1.19$\pm$0.03 & 0.54 & -1.00$\pm$0.06 &       & $<$10 ($<$80.1) & $>$1.0 & $>$25.33  \\
GMRT022455-032906 & 20.7$\pm$0.6 & 12.9$\pm$0.17 & 0.85$\pm$0.15 &         & -0.61$\pm$0.04 & -1.71$\pm$0.10    & 1.10 & -1.33$\pm$0.06 &       & 16 ($<$128.1) & $>$1.0  &  $>$26.13  \\
GMRT022611-050508 & 18.8$\pm$0.7  & 11.2$\pm$0.16 & 1.08$\pm$0.01 &        & -0.67$\pm$0.05 & -1.60$\pm$0.01 & 0.93 & -1.28$\pm$0.02 &       & $<$10 ($<$80.1) & $>$1.0  & $>$26.08 \\
GMRT022627-033301 & 18.4$\pm$0.7  & 11.37$\pm$0.21& 0.62$\pm$0.15 &        & -0.62$\pm$0.05 & -1.65$\pm$0.10 & 1.37 & -1.29$\pm$0.07 &       & $<$10 ($<$83.4) & 1.253$\pm$0.126  &  26.37  \\
GMRT022631-043926 & 14.2$\pm$0.4  & 9.11$\pm$0.38 & 1.32$\pm$0.01 &        & -0.57$\pm$0.06 & -1.32$\pm$0.03 & 0.75 & -1.06$\pm$0.01 &       & 18.4 (27.1) &  0.0778$\pm$0.006  &  23.33  \\
GMRT022654-025821 & 7.1$\pm$0.7   & 4.67$\pm$0.35 & $<$1.00  &           & -0.53$\pm$0.16 &$<$-1.05     & $>$0.52 & $<$-0.87  &        & $<$10 ($<$80.1) & $>$1.0   &  $>$25.53 \\
GMRT022713-031247 & 7.9$\pm$0.8   & 5.02$\pm$0.27 & $<$1.00  &            & -0.58$\pm$0.14 &$<$-1.10     & $>$0.52 & $<$-0.92 &        & $<$10 ($<$79.7) & 0.979$\pm$0.0848  &  25.57  \\
GMRT022717-061832 & 9.9$\pm$0.8   & 6.16$\pm$0.42 & $<$1.00  &    & -0.63$\pm$0.14 &$<$-1.25 & $>$0.62 & $<$-1.03  &      & $<$10 ($<$46.3) & 0.317$\pm$0.029  & 24.52  \\
GMRT022720-033039 & 28.7$\pm$1.0  & 16.4$\pm$0.27 & 2.30$\pm$0.5 &        & -0.72$\pm$0.05 & -1.35$\pm$0.15 & 0.62 & -1.13$\pm$0.10 &       & 21.0 (168.2) & $>$1.0  & $>$26.22  \\
GMRT022723-051242 & 47.1$\pm$0.8  & 25.9$\pm$0.63 & 3.20$\pm$0.5 &         & -0.77$\pm$0.04 & -1.43$\pm$0.11 & 0.66 & -1.20$\pm$0.07 &       & 27 (216.3) & $>$1.0   &  $>$26.45  \\
GMRT022737-052139 & 6.3$\pm$0.8   & 3.99$\pm$0.20 & 0.75$\pm$0.01 &        & -0.58$\pm$0.17 & -1.14$\pm$0.04 & 0.56 & -0.94$\pm$0.05 &       & 19.5 (156.2) &  $>$1.0  &  $>$25.49 \\
GMRT022750-024524 & 12.1$\pm$0.9  & 7.46$\pm$0.65 & $<$1.0 &            & -0.63$\pm$0.15 &$<$-1.38        & $>$0.79 & 
$<$-1.12  &       & $<$10 ($<$82.8) & 1.185  & 26.02 \\
GMRT022759-051354 & 4.1$\pm$0.8   & 2.73$\pm$0.16 & 0.58$\pm$0.01 &        & -0.52$\pm$0.25 & -1.05$\pm$0.04 & 0.53 & -0.86$\pm$0.08 &       & $<$10 ($<$80.1) & $>$1.0  & $>$25.29  \\
GMRT022802-041417 & 37.3$\pm$0.9  & 22.06$\pm$0.7 & 3.20$\pm$0.5 &        & -0.68$\pm$0.05 & -1.32$\pm$0.11 & 0.64 & -1.09$\pm$0.07 &       & 19.1 (112.3) & 0.467$\pm$0.0344 &  25.49  \\
\hline
\end{tabular}}
\\
Notes - An upper limit of 10$^{\prime\prime}$ is kept for the size of unresolved sources.  
\end{table} 
\begin{figure*}
\includegraphics[angle=0,width=5.8cm,trim={1.0cm 2.0cm 3.5cm 0.5cm},clip]{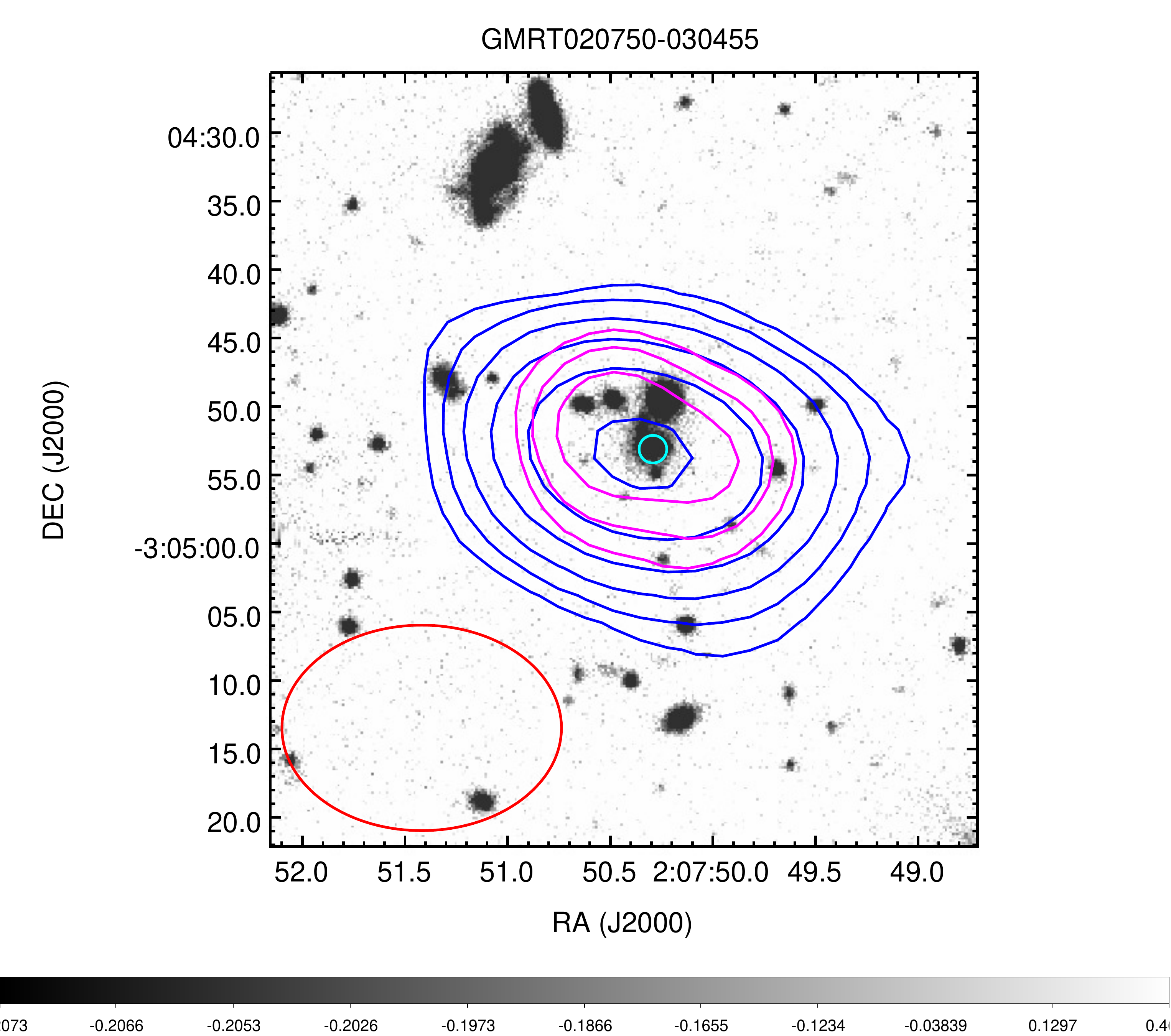}
\includegraphics[angle=0,width=6.2cm,trim={0.5cm 2.0cm 3.0cm 2.0cm},clip]{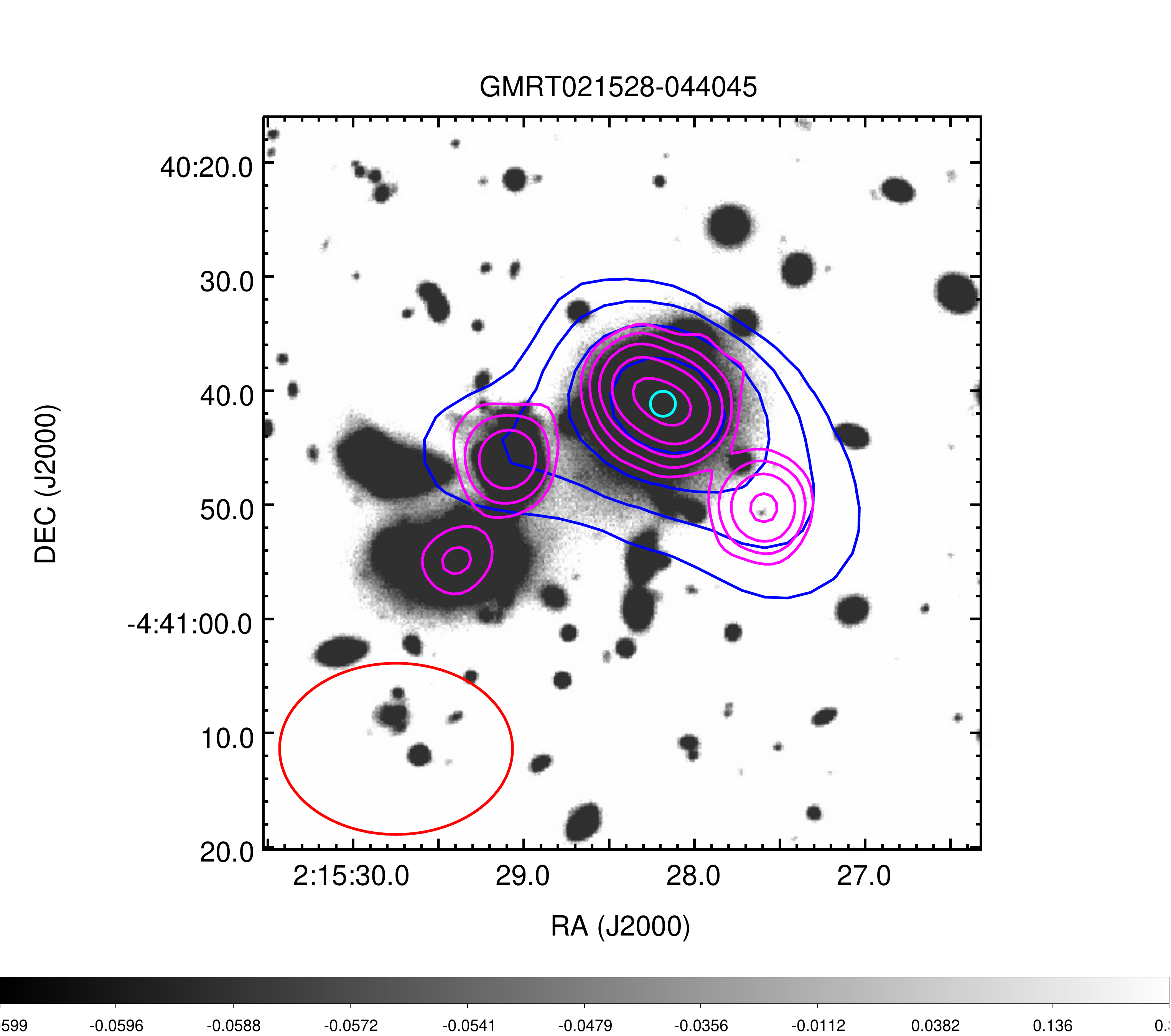}
\includegraphics[angle=0,width=6.2cm,trim={1.0cm 2.0cm 3.0cm 1.5cm},clip]{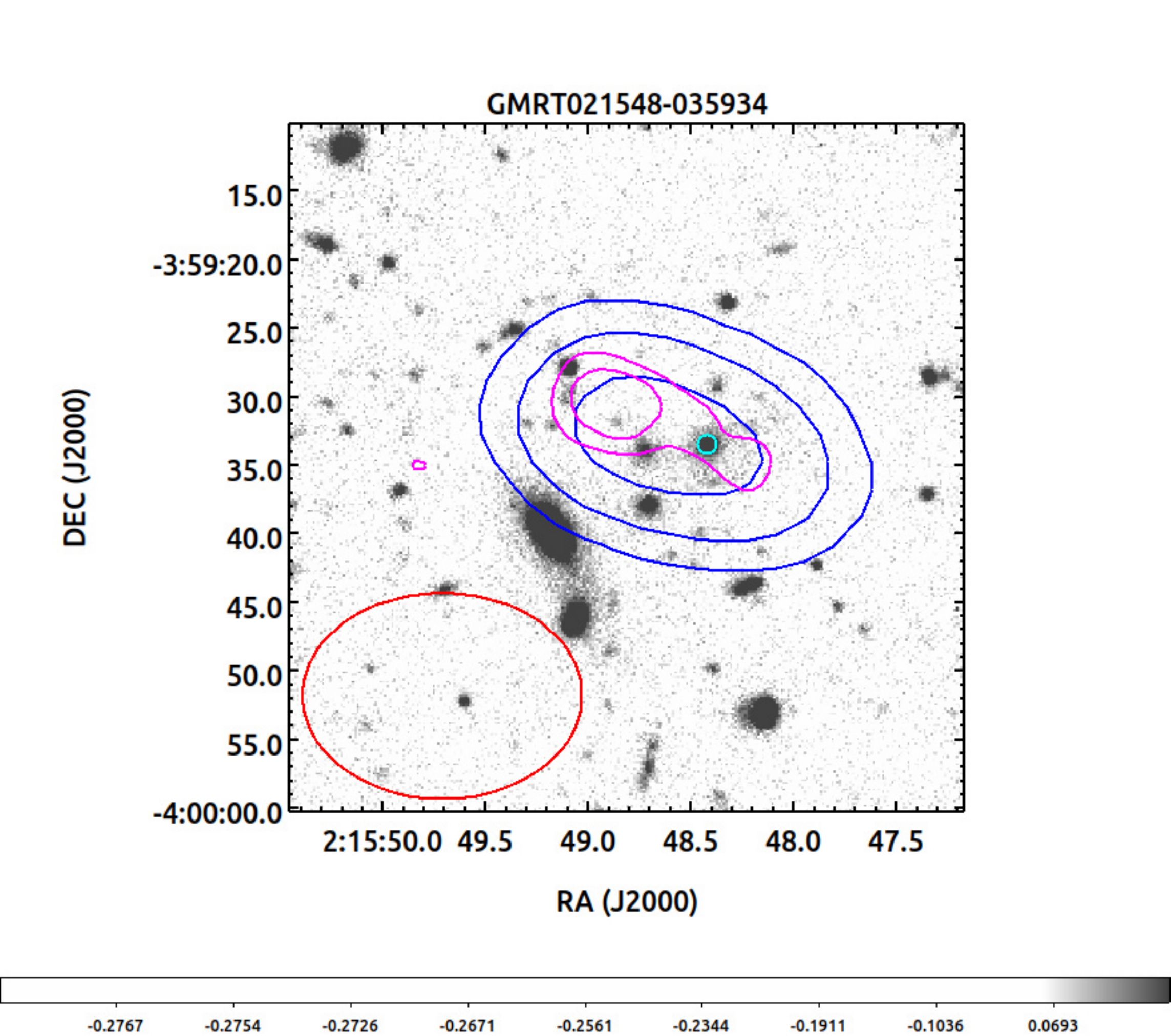}
{\includegraphics[angle=0,width=6.0cm,trim={2.5cm 2.5cm 10.0cm 3.0cm},clip]{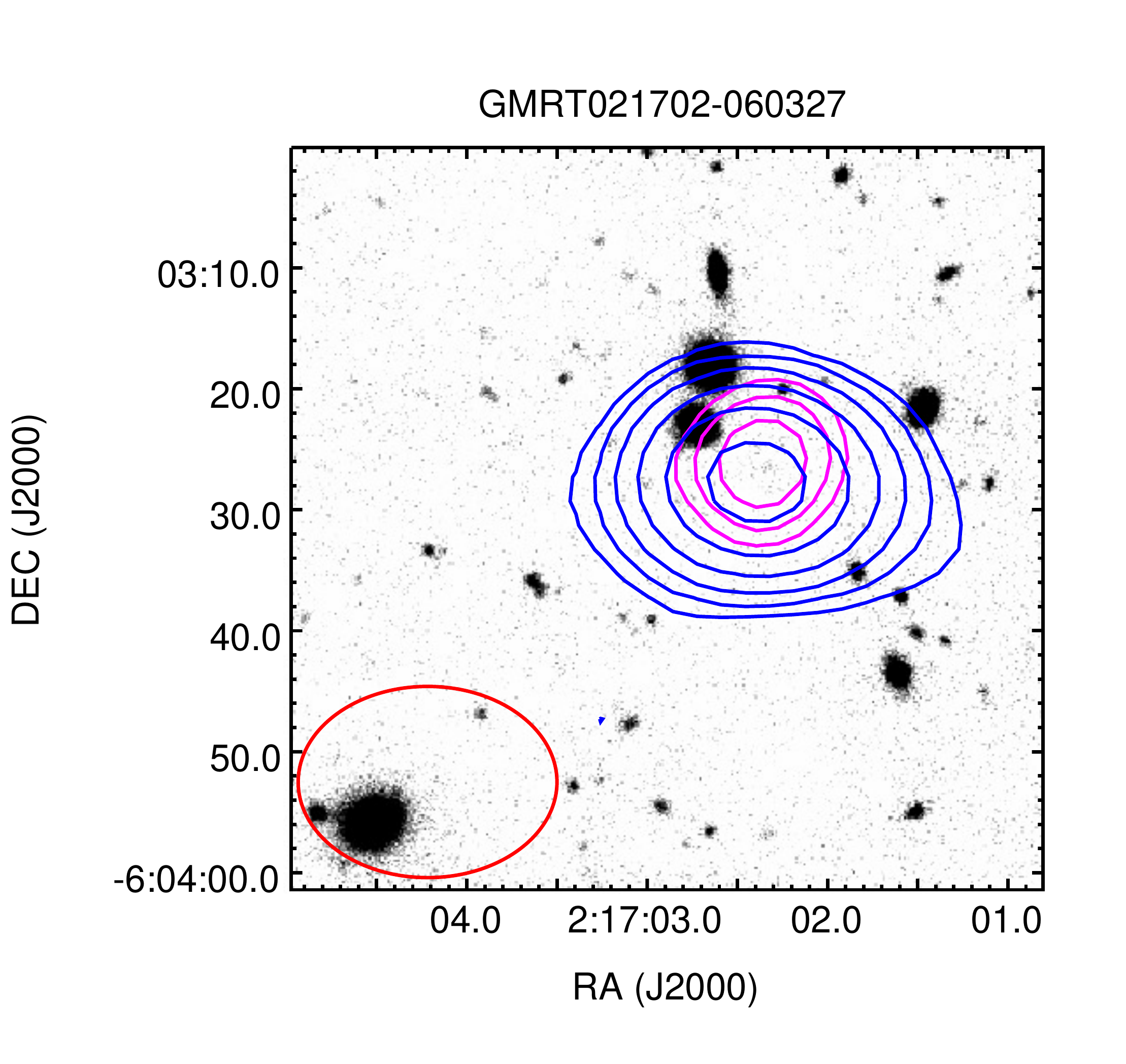}}
{\includegraphics[angle=0,width=6.0cm,trim={1.0cm 2.0cm 7.0cm 1.5cm},clip]{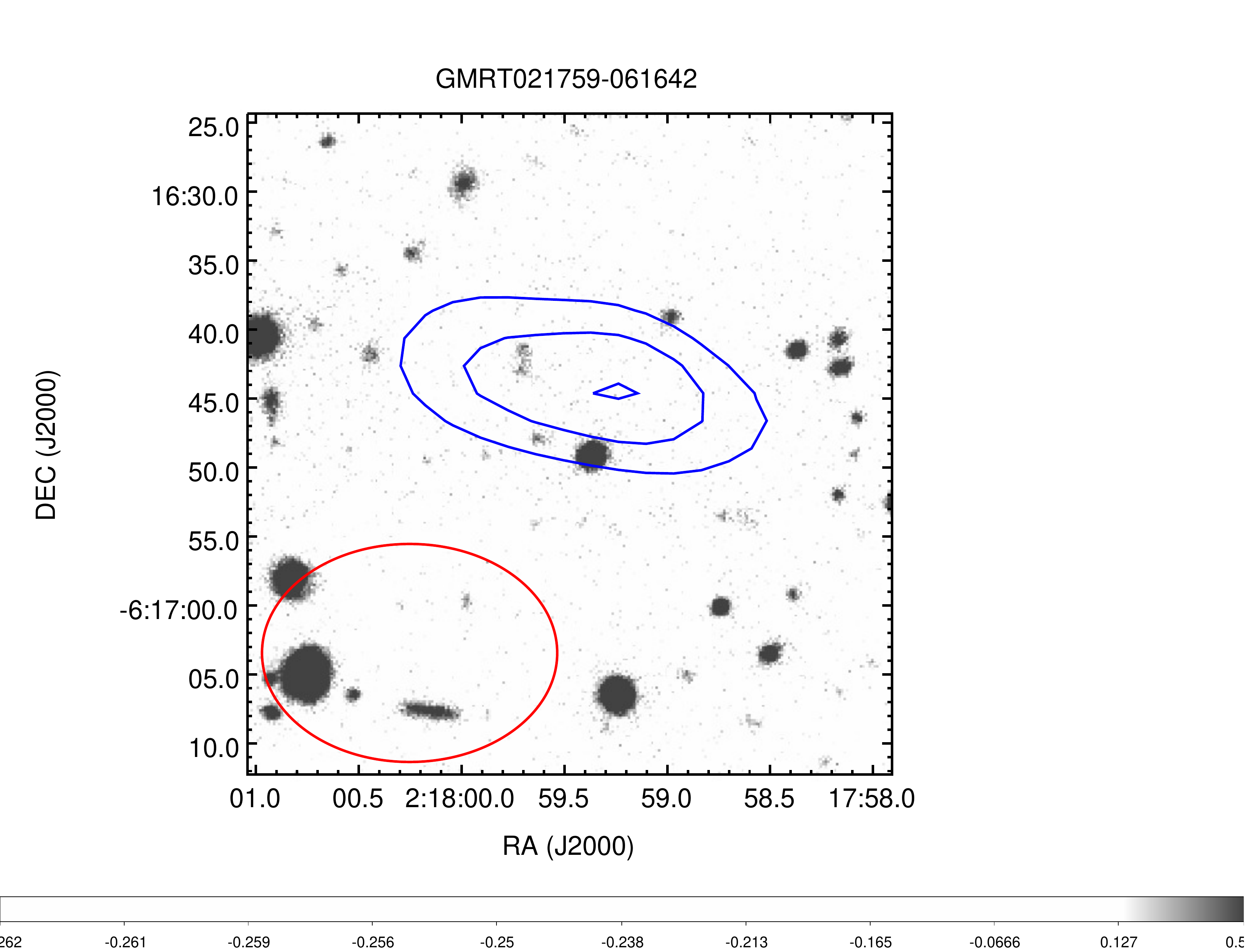}}
{\includegraphics[angle=0,width=6.0cm,trim={2.0cm 2.5cm 7.0cm 1.5cm},clip]{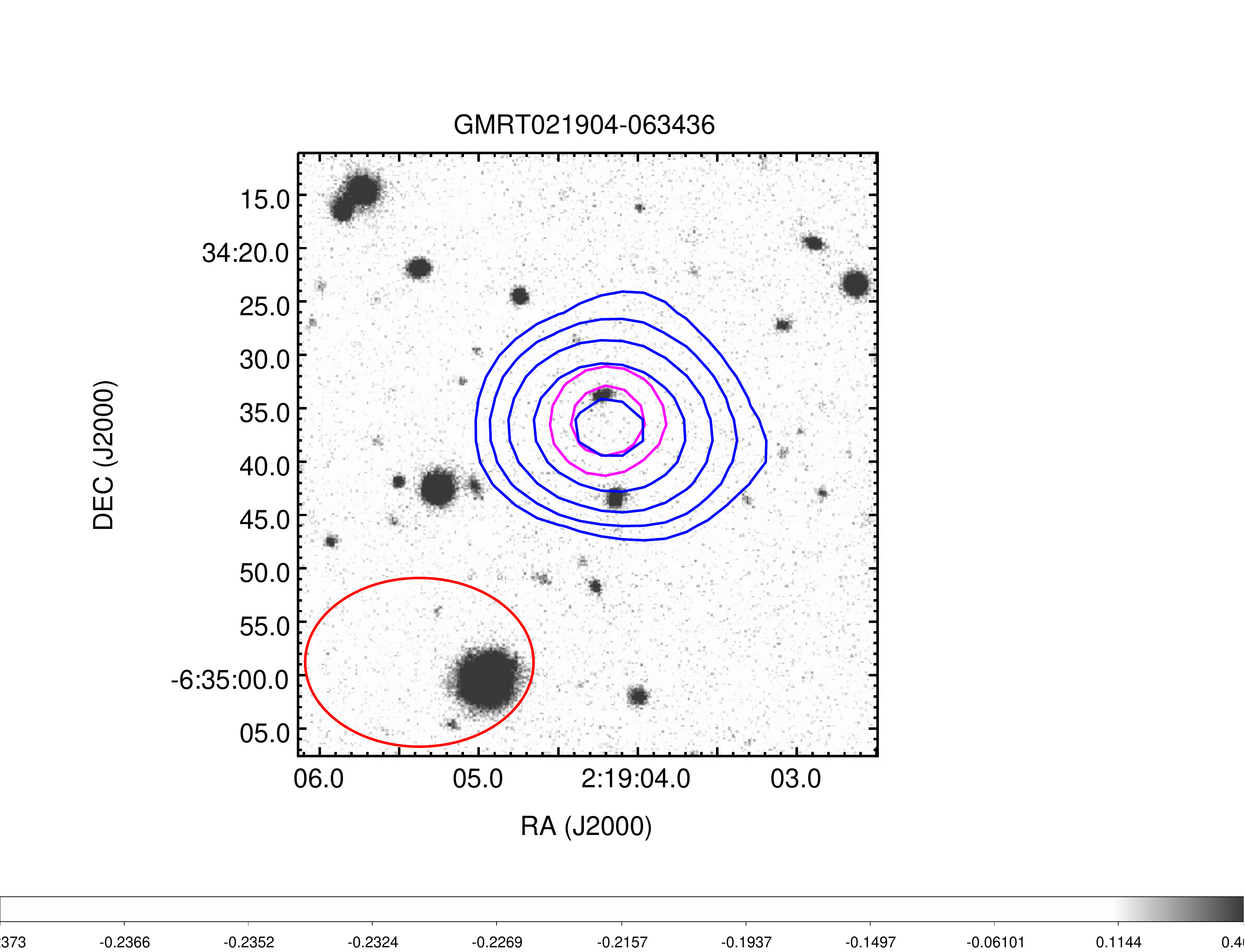}}
\includegraphics[angle=0,width=6.0cm,trim={1.0cm 2.0cm 4.0cm 2.5cm},clip]{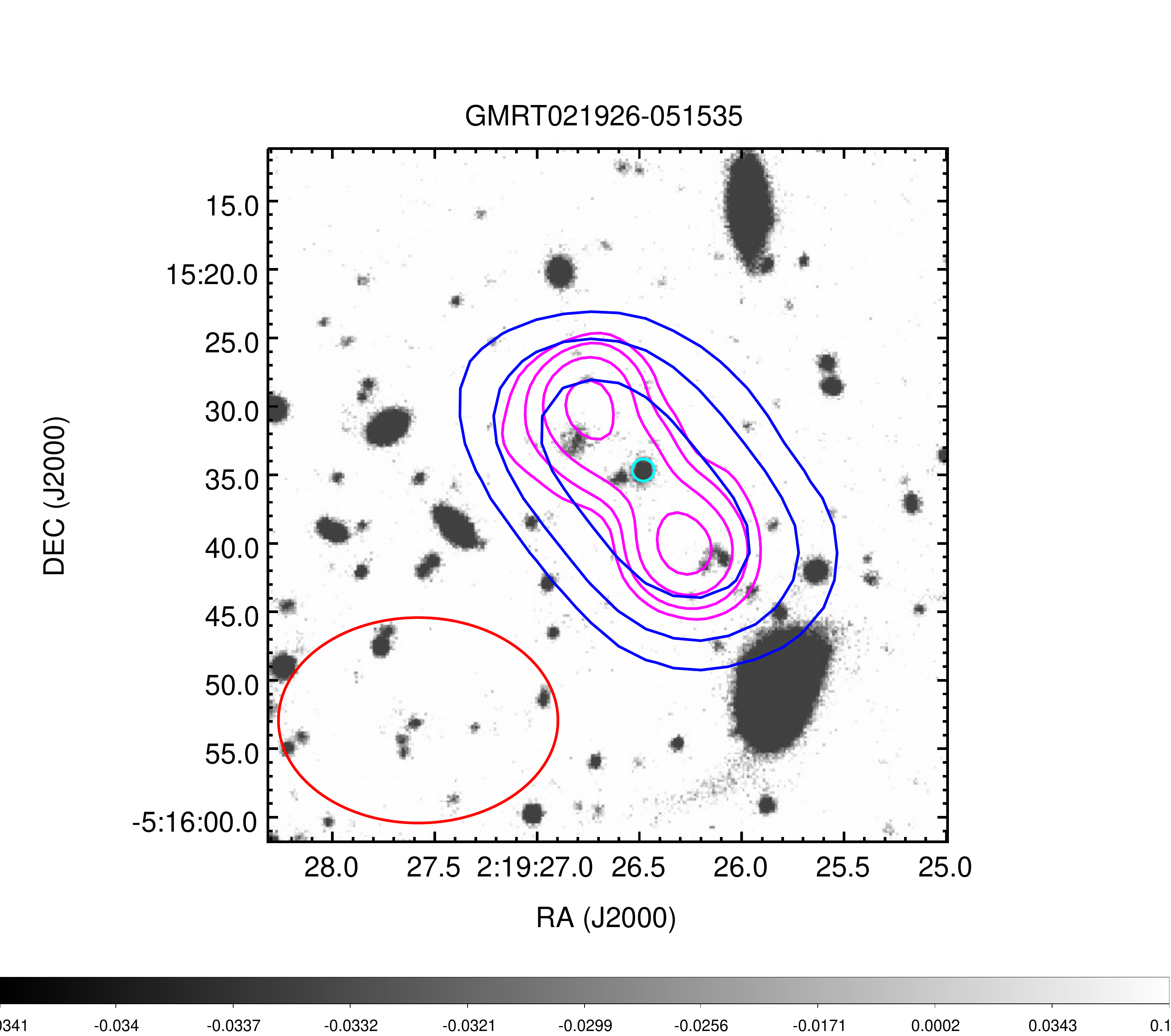}
\includegraphics[angle=0,width=6.0cm,trim={1.0cm 1.75cm 8.5cm 3.0cm},clip]{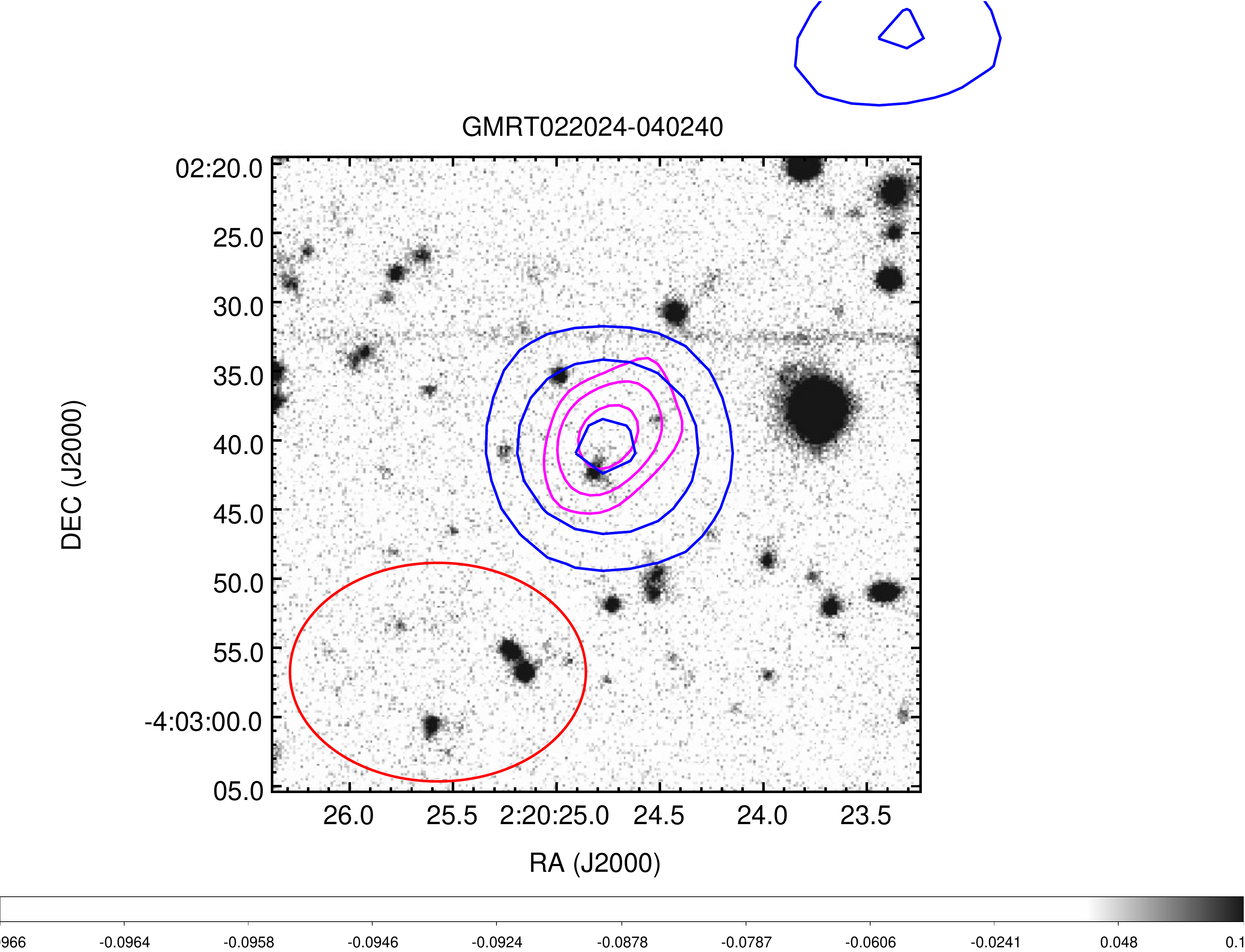}
\includegraphics[angle=0,width=6.0cm,trim={1.0cm 2.5cm 4.0cm 2.5cm},clip]{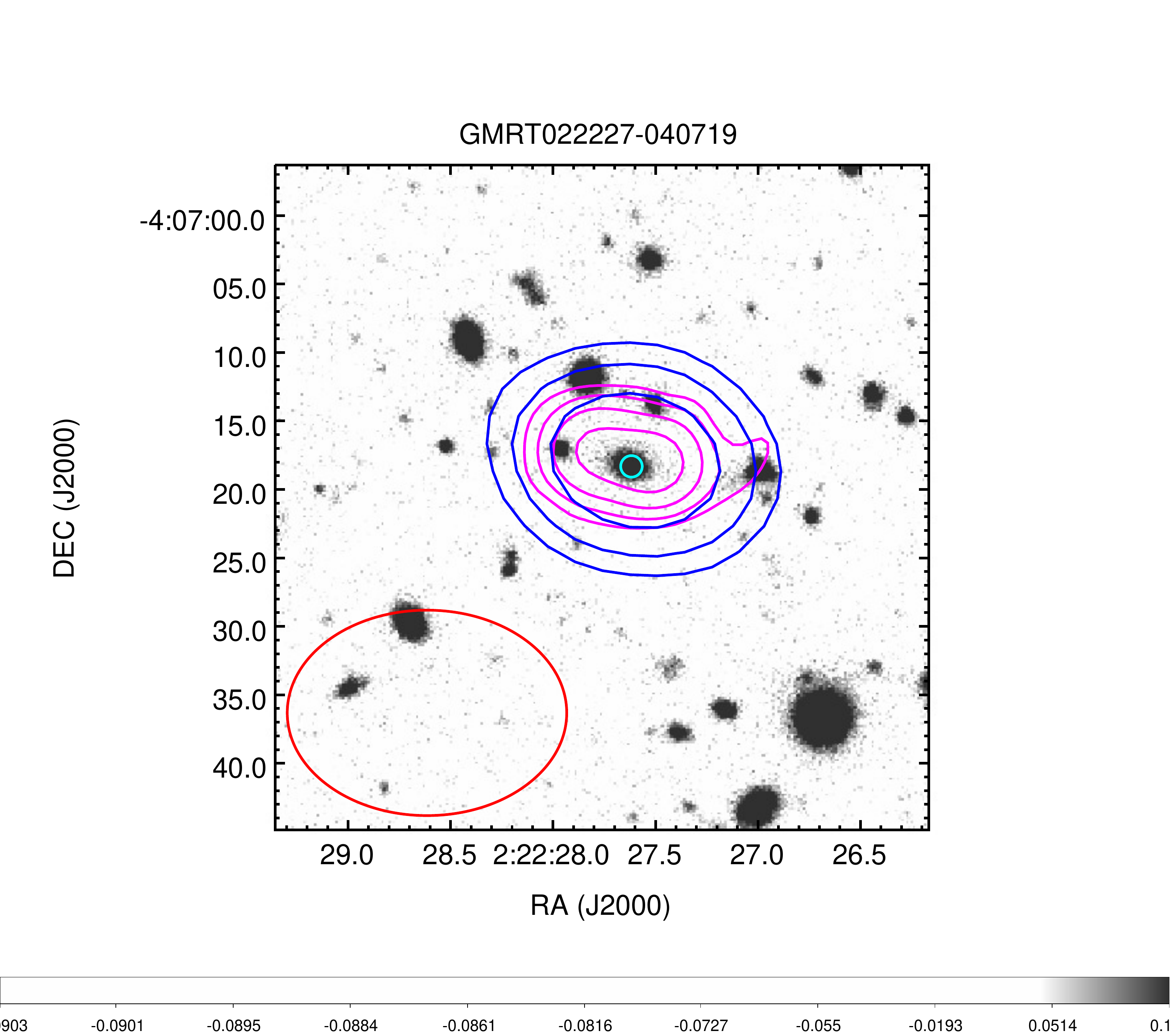}
{\includegraphics[angle=0,width=6.2cm,trim={1.0cm 2.0cm 2.0cm 1.0cm},clip]{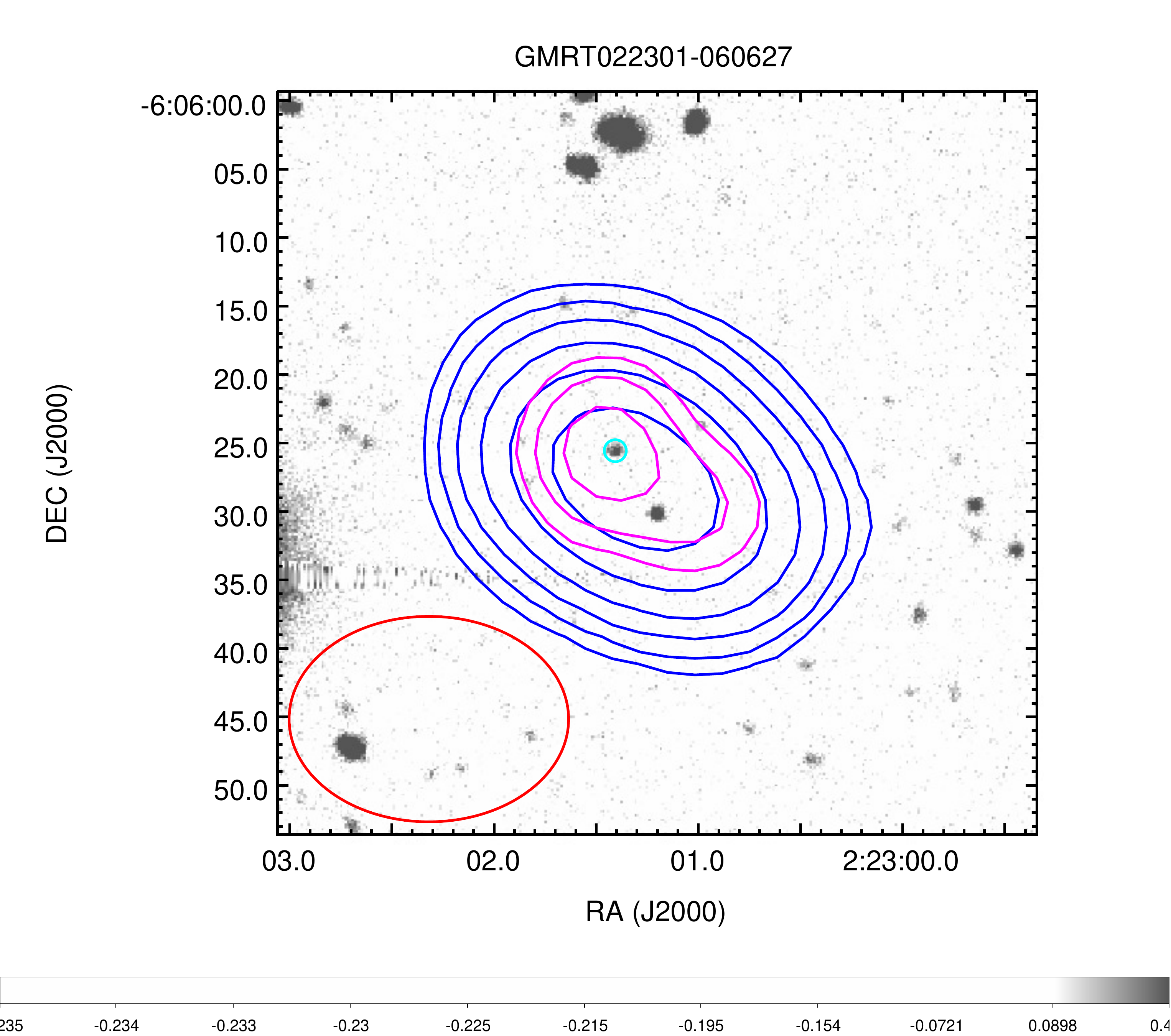}}
{\includegraphics[angle=0,width=6.0cm,trim={1.5cm 2.0cm 9.0cm 1.0cm},clip]{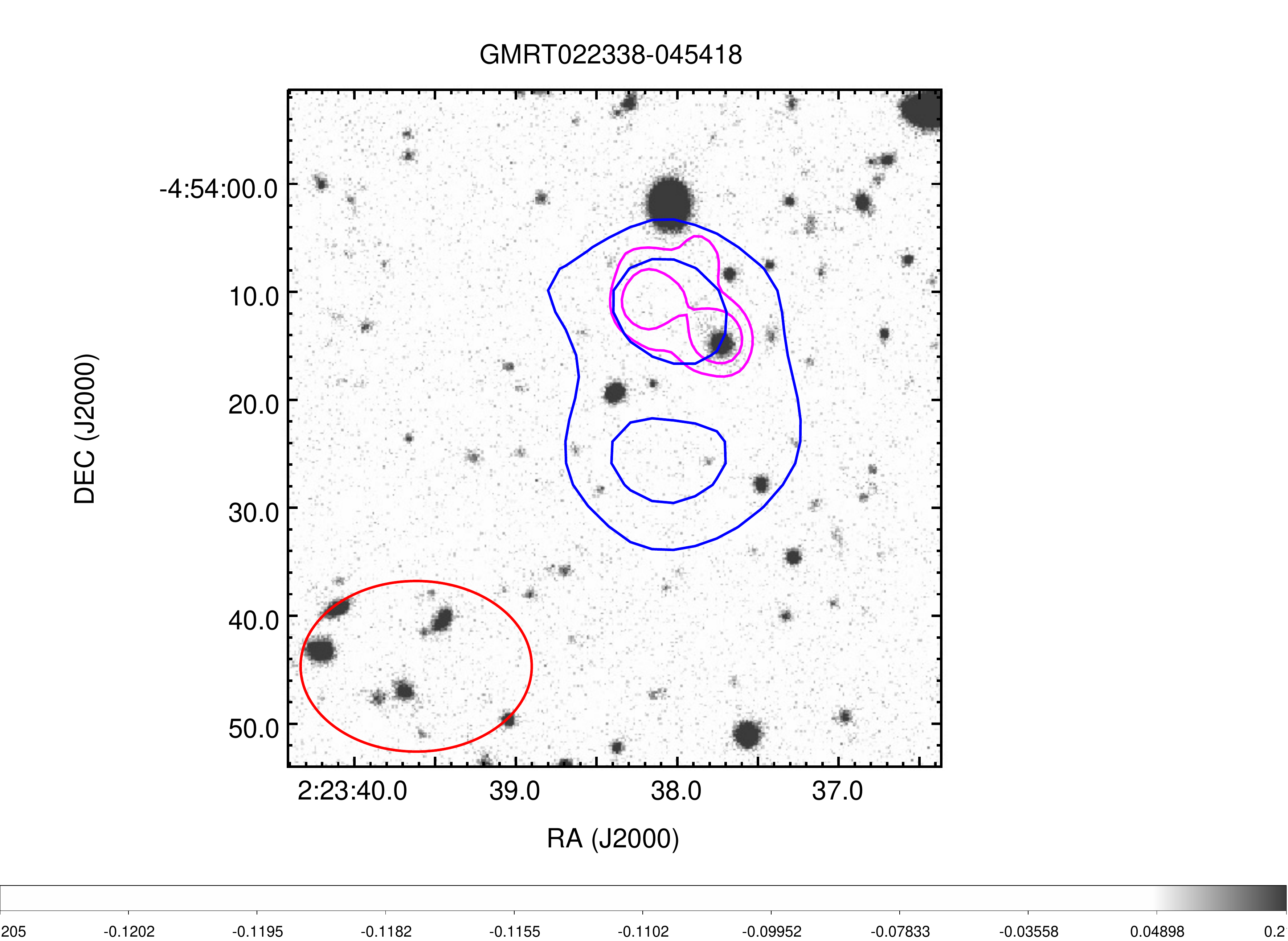}}
{\includegraphics[angle=0,width=6.0cm,trim={0.25cm 1.65cm 2.5cm 0.5cm},clip]{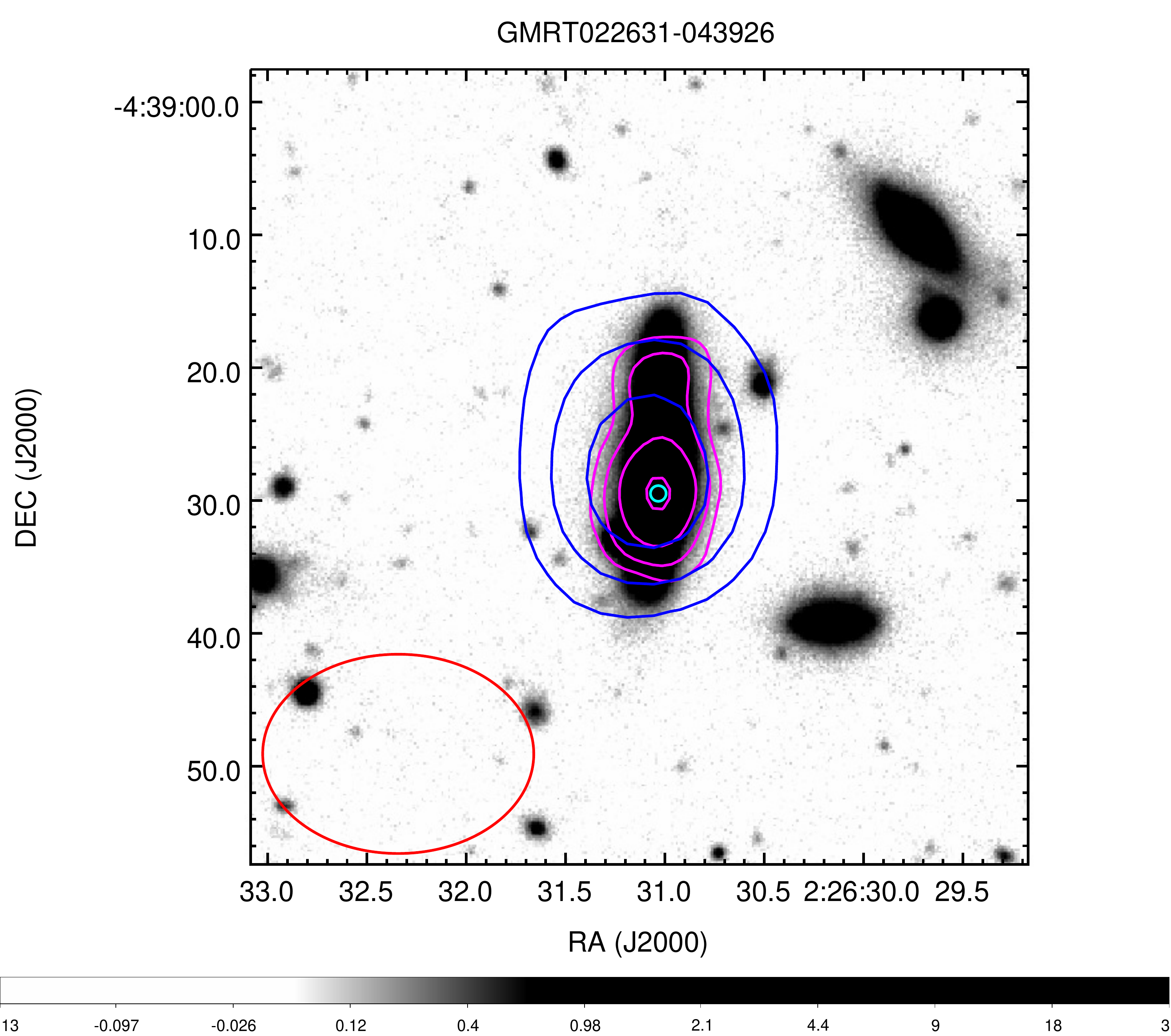}}
\caption{Images of remnant candidates showing extended radio emission. 
The 325 MHz GMRT radio contours (in Blue) and 1.4 GHz JVLA radio contours (in Magenta) are overlaid on the corresponding 
$i$ band HSC-SSP optical images. Radio contours are at 3$\sigma$ × (1, 2, 4, 8, 16 ......) levels and 
the corresponding optical image is logrithmically scaled. Potential host galaxy is marked with a small circle (in Cyan) around it. 
The 325 MHz GMRT synthesized beam of 10$^{\prime\prime}$.7 $\times$ 7$^{\prime\prime}$.9 is shown 
by an ellipse (in Red) in the bottom left corner in each plot. The HSC-SSP deep survey images, whenever available, 
are preferred over the HSC-SSP wide survey.}
\label{fig:RadioCont} 
\end{figure*}
\addtocounter{figure}{-1}
\begin{figure*}
\includegraphics[angle=0,width=6.0cm,trim={1.0cm 2.0cm 3.0cm 1.0cm},clip]{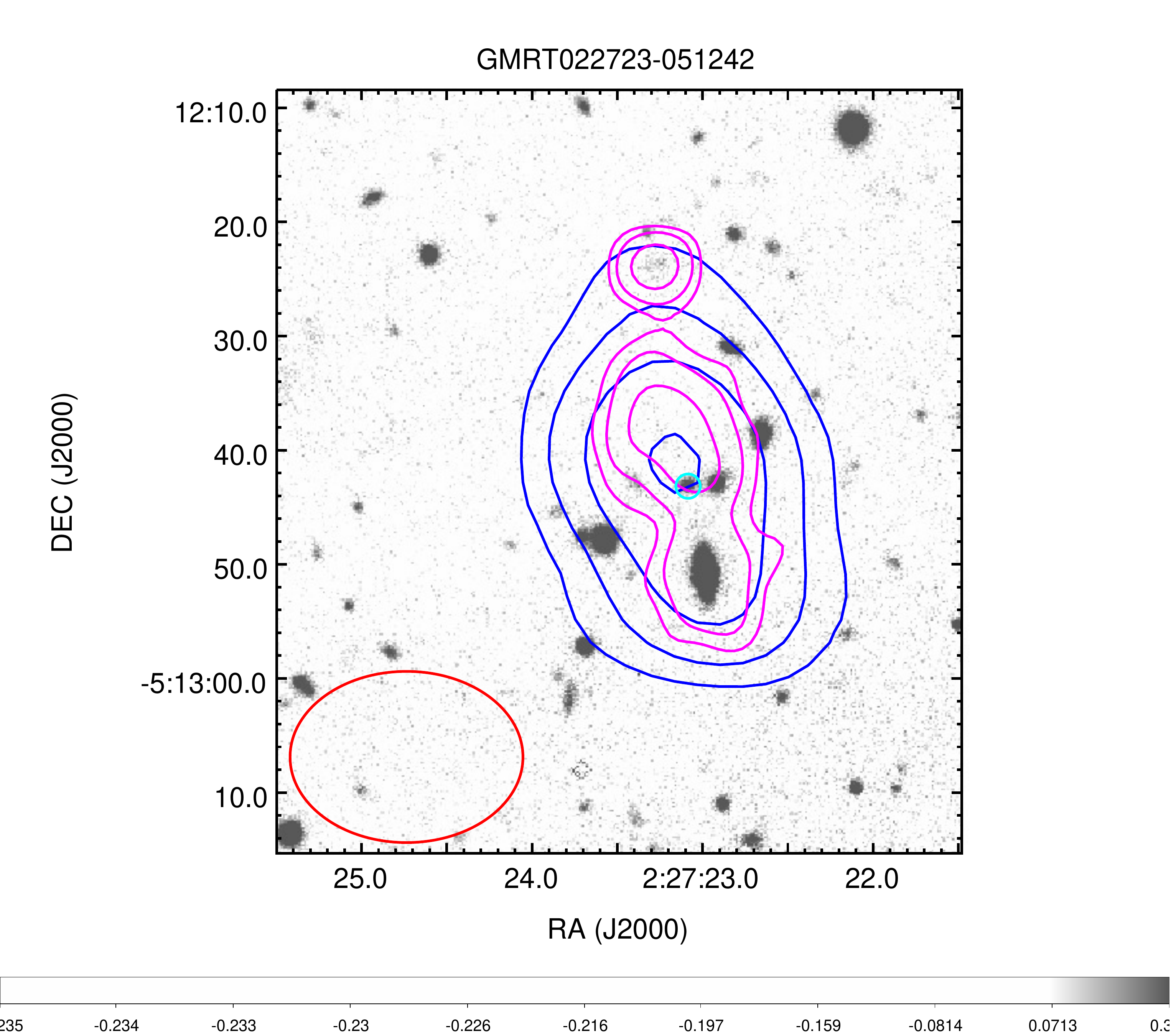}
\includegraphics[angle=0,width=6.0cm,trim={0.5cm 2.0cm 3.0cm 0.0cm},clip]{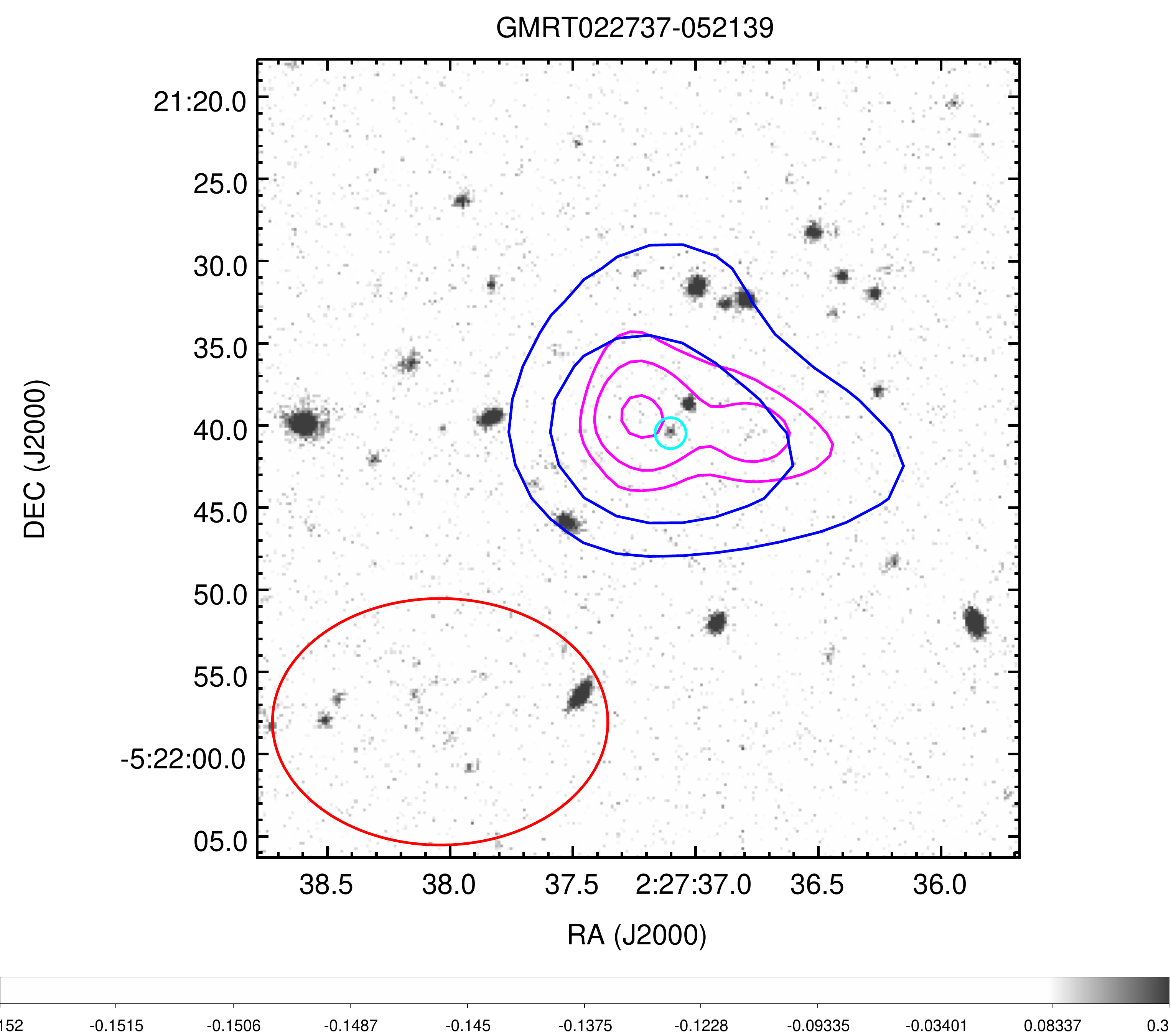}
\includegraphics[angle=0,width=6.0cm,trim={1.0cm 1.5cm 3.5cm 1.5cm},clip]{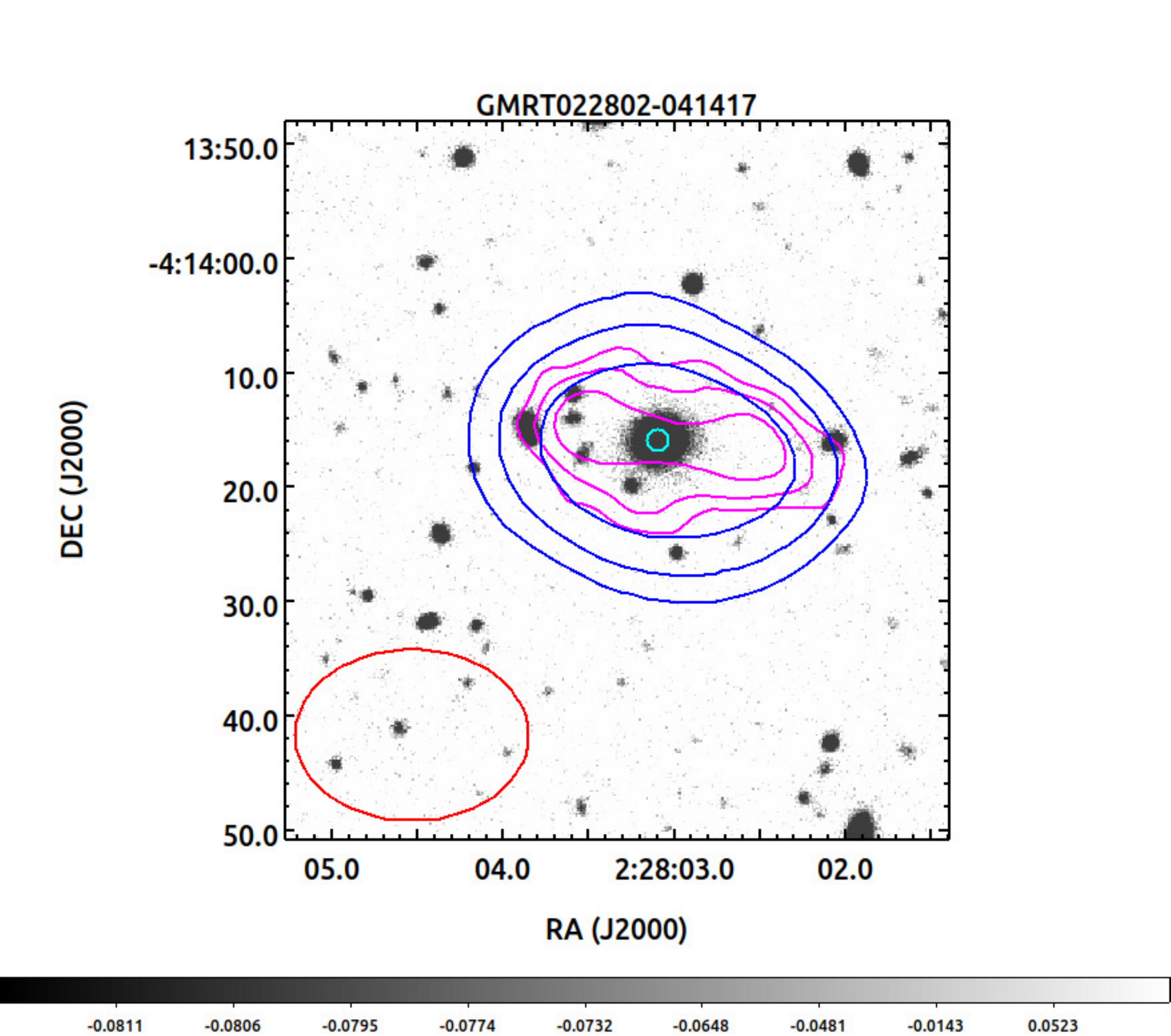}
\caption{{\it Continue}.}
\label{fig:RadioCont} 
\end{figure*}
\begin{figure*}
\includegraphics[angle=0,width=6.0cm,trim={2.0cm 1.75cm 7.0cm 2.0cm},clip]{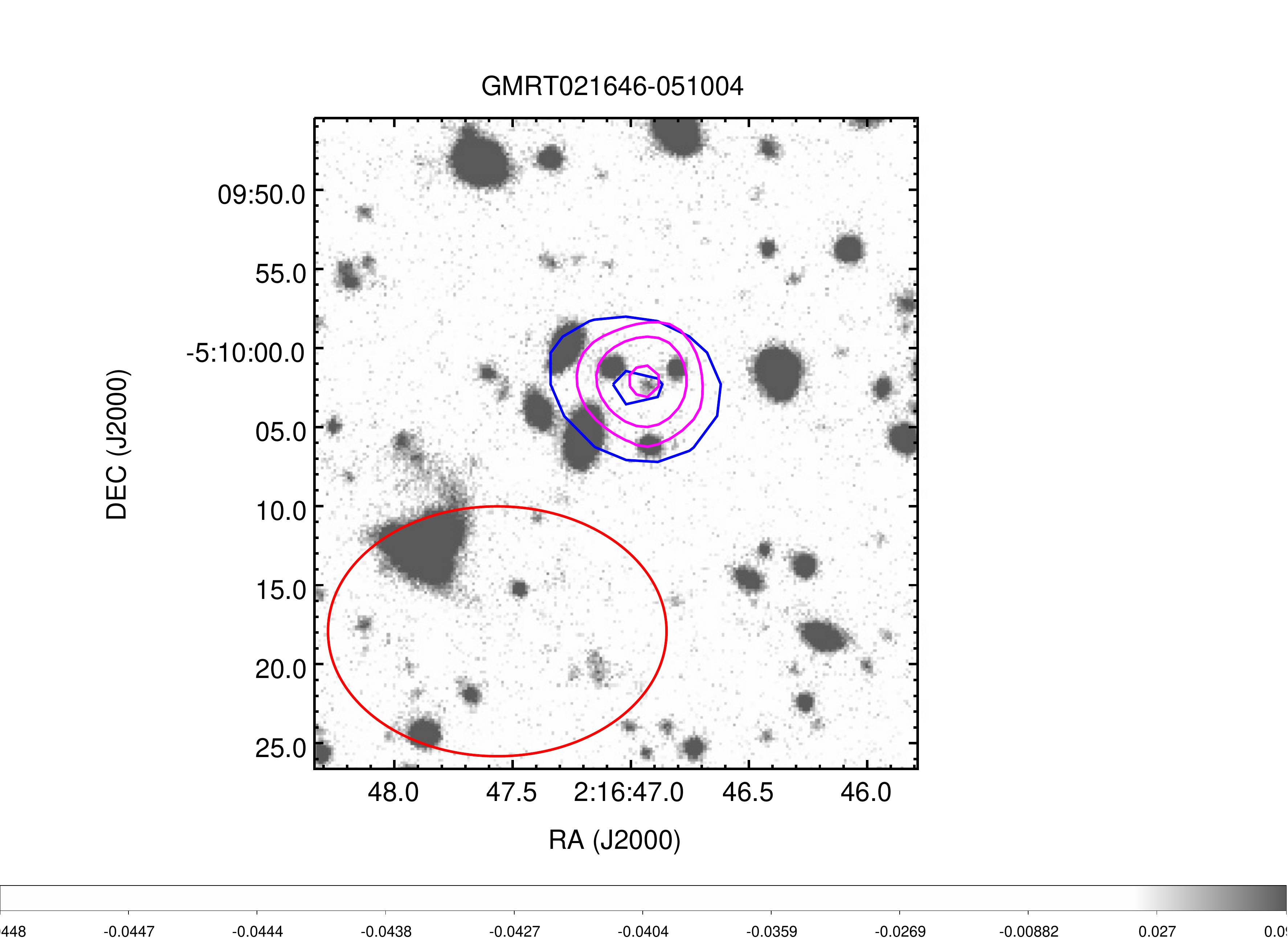}
\includegraphics[angle=0,width=6.0cm,trim={1.5cm 1.65cm 6.0cm 1.0cm},clip]{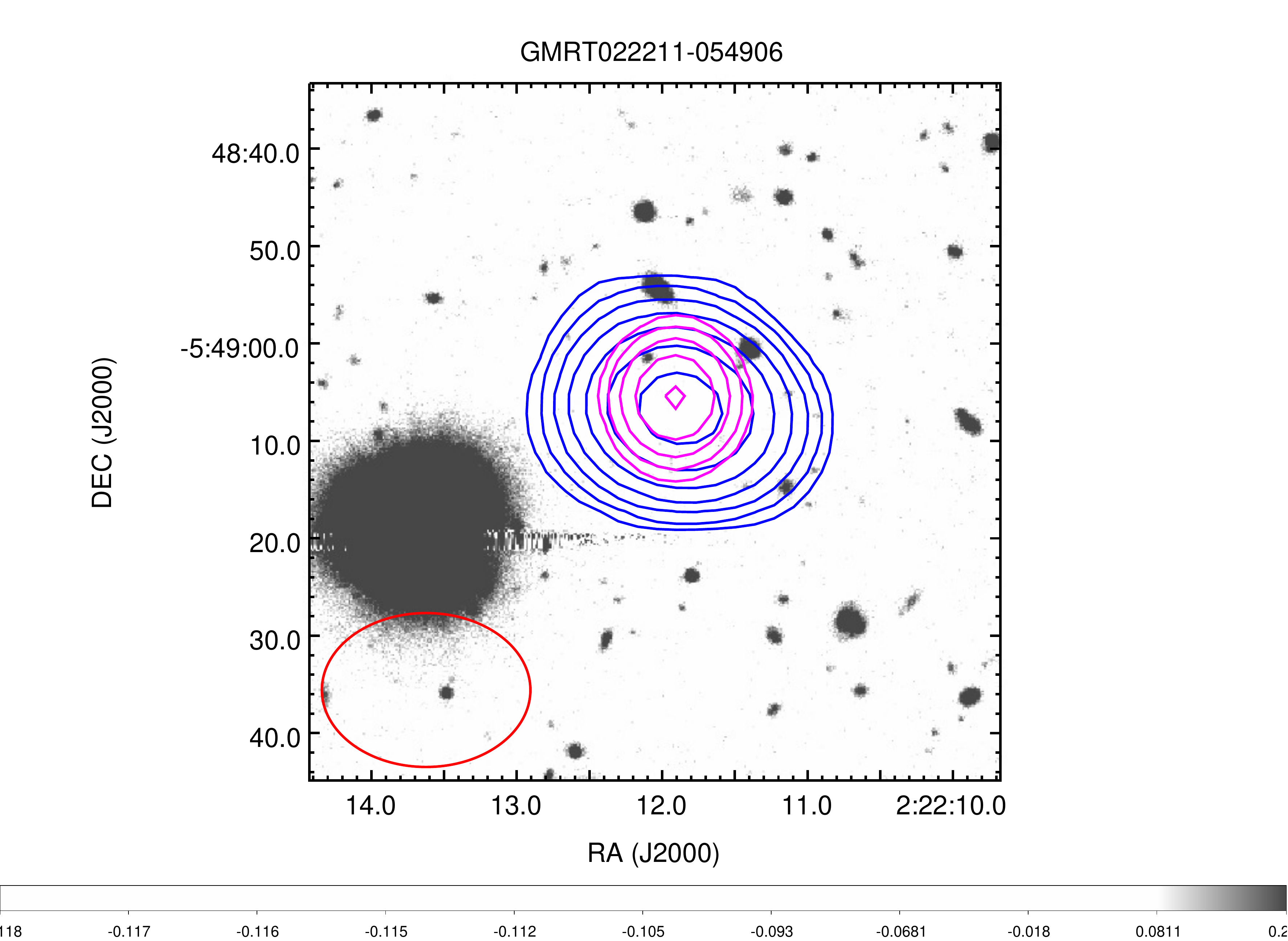}
\includegraphics[angle=0,width=6.0cm,trim={2.0cm 1.75cm 7.0cm 1.5cm},clip]{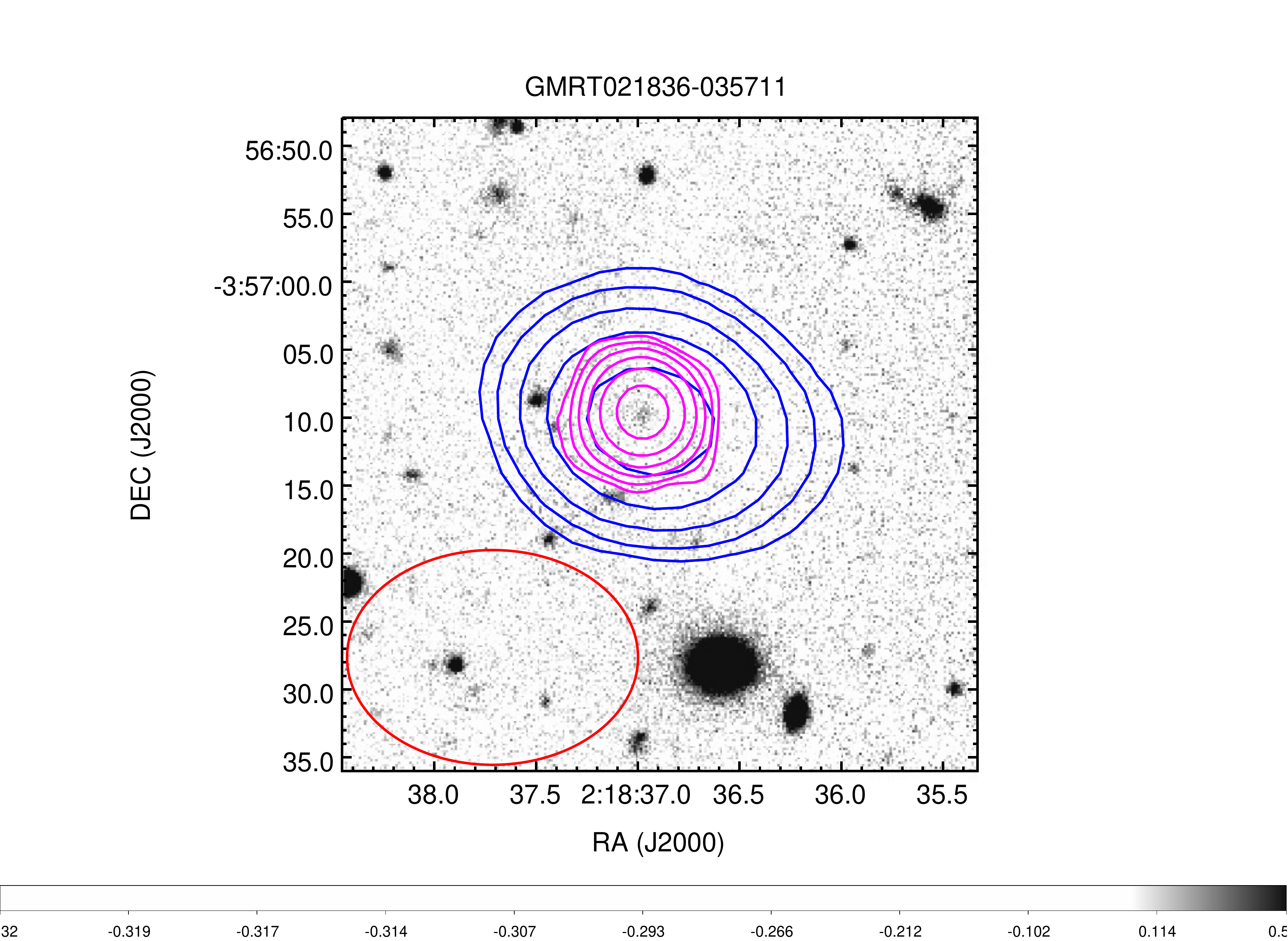}
\caption{Three examples of unresolved remnant candidates. {\it Left panel}: a faint source with S$_{\rm 325~MHz}$ = 1.93$\pm$0.12 mJy. 
{\it Middle panel}: a source with an intermediate level flux density of S$_{\rm 325~MHz}$ = 17.4$\pm$0.2 mJy. 
{\it Right panel}: a bright source with S$_{\rm 325~MHz}$ = 121.4$\pm$0.3 mJy.}
\label{fig:RadioContUR} 
\end{figure*}
\begin{figure*}
\includegraphics[angle=0,width=6.0cm,trim={0.0cm 0.0cm 0.0cm 0.0cm},clip]{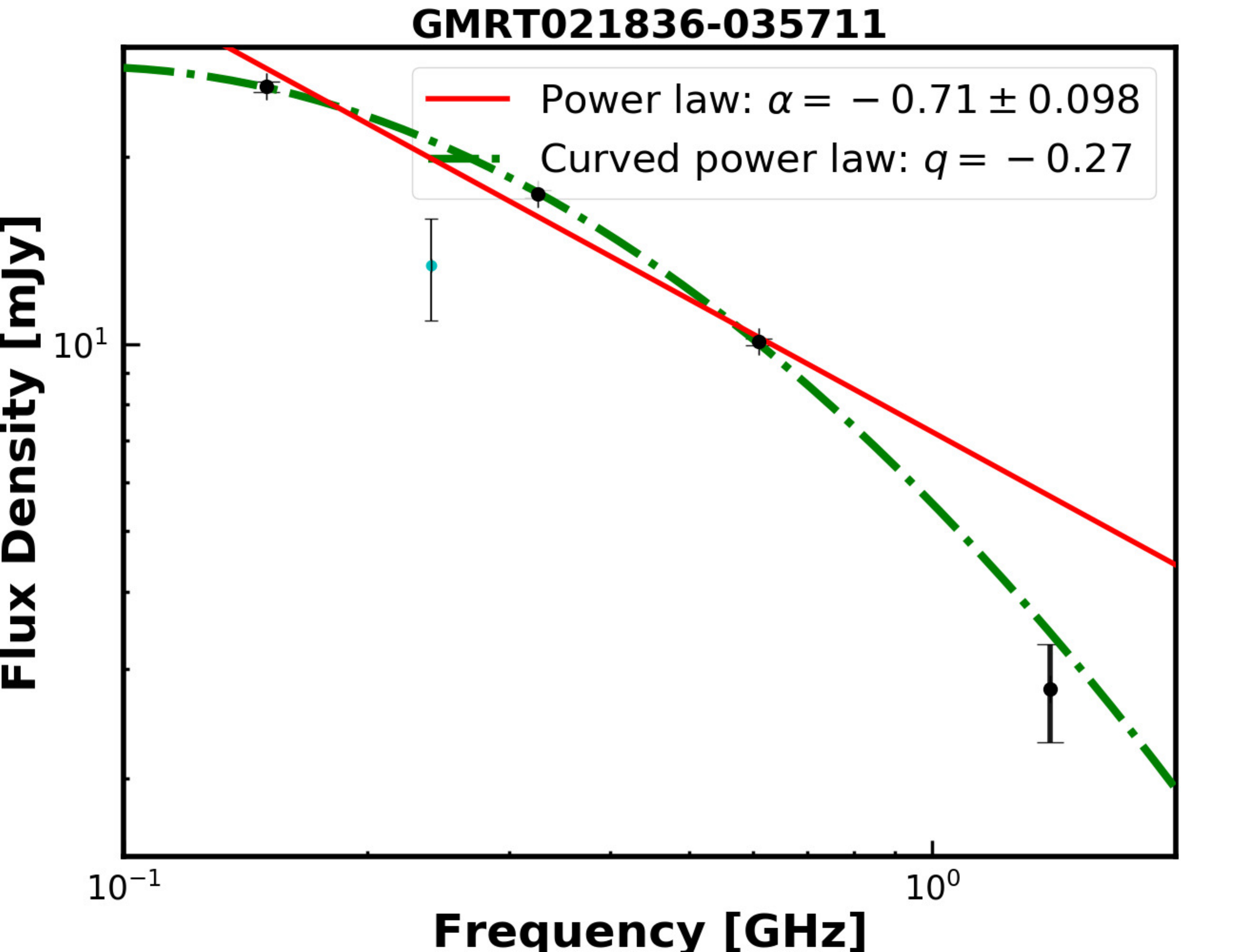}
\includegraphics[angle=0,width=6.0cm,trim={0.0cm 0.0cm 0.0cm 0.0cm},clip]{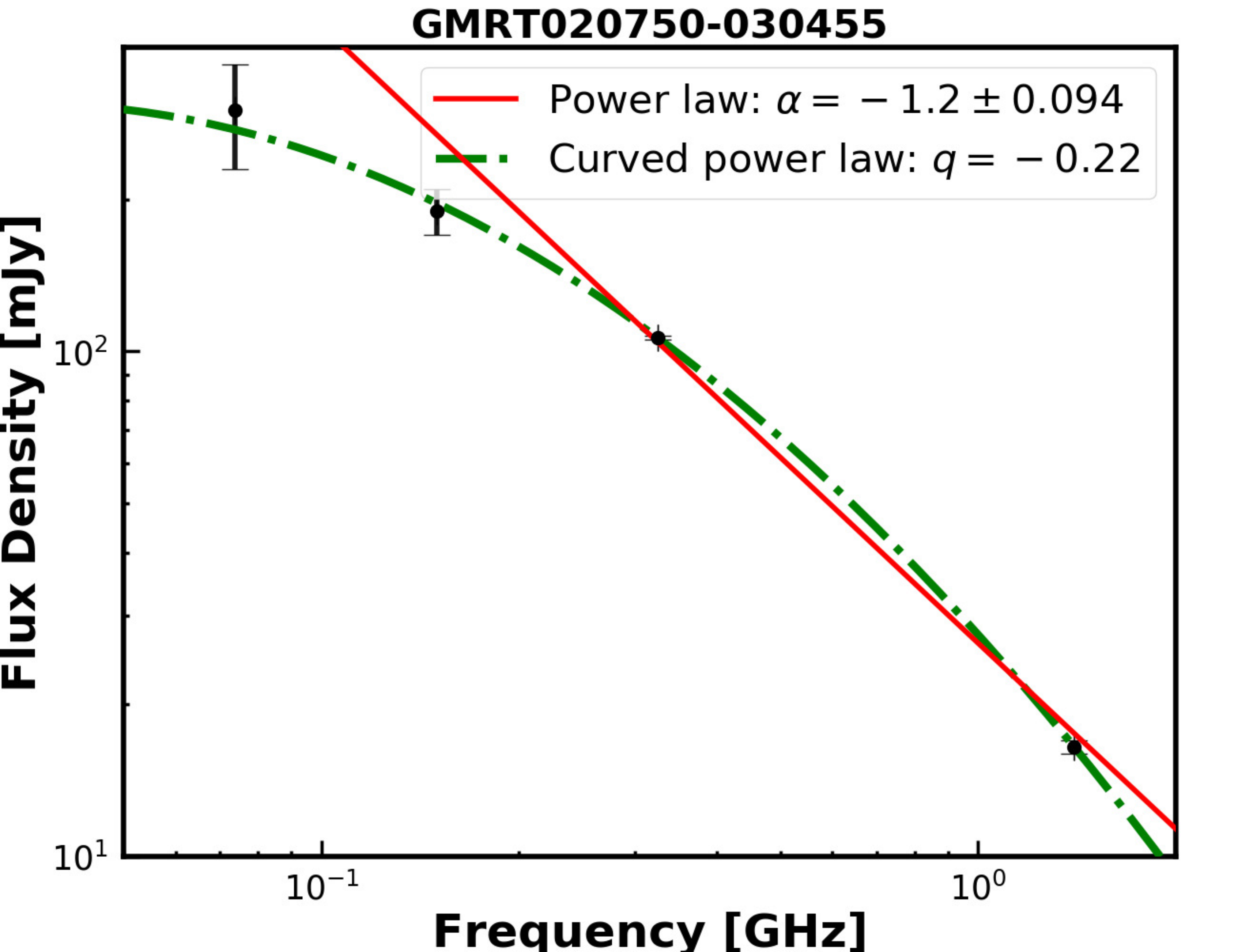}
\includegraphics[angle=0,width=6.0cm,trim={0.0cm 0.0cm 0.0cm 0.0cm},clip]{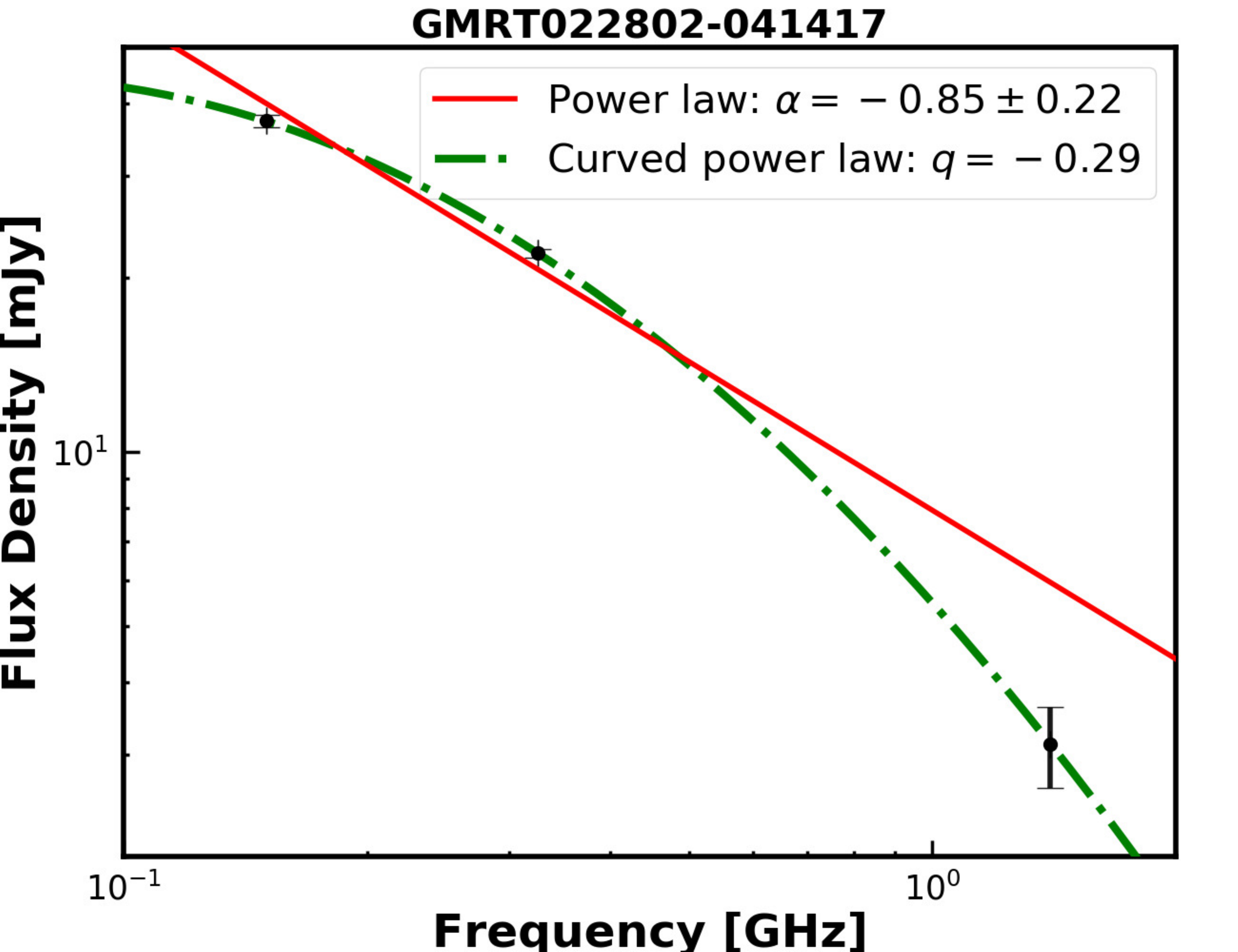}
\caption{Three examples of radio spectra showing strong spectral curvature. Spectra are fitted with the curved 
power law and power law models. {\it Left panel} : GMRT 021836-035711 $-$ an unresolved remnant candidate. 
{\it Middle panel}: GMRT020750-030455 $-$ a remnant candidate showing slightly resolved emission. 
{\it Right panel}: GMRT022802-041417 $-$ a remnant candidate showing double-lobe morphology in the 1.4 GHz JVLA image.}
\label{fig:Spectra} 
\end{figure*}
%
%
\subsection{Radio morphology}
\label{sec:morph}
Our remnant candidates are selected to be of small-sizes (LAS $\leq$ 30$^{\prime\prime}$), 
and hence, they are not well suited for revealing the morphological details.  
Despite the limitations imposed by small size we examine radio morphologies of our remnant candidates 
by inspecting 325 MHz GMRT and 1.4 GHz radio images. We use 1.4 GHz images either from the JVLA survey, whenever available, 
or from the FIRST survey due to their relatively higher resolution of 5$^{\prime\prime}$.0. 
We classify a source as resolved if extended emission is seen either in the 1.4 GHz image or in the 325 MHz image. 
We find that only 15/48 (31$\%$) of our remnant candidates show resolved emission {\ie}emission at scales 
larger than the synthesized beam. As expected morphological features are often seen only in the 1.4 GHz images 
due to its higher resolution. 
In Figure~\ref{fig:RadioCont} we show radio contours (325 MHz (in Blue) and 1.4 GHz (in Magenta)) over-plotted onto the 
$i$ band HSC-SSP optical images for all 15 remnant candidates. In several sources 1.4 GHz radio contours 
clearly show double-lobe radio morphology, for example, GMRT021926-051535, GMRT022723-051242, 
and GMRT022802-041417 (see Figure~\ref{fig:RadioCont}). In a few cases (GMRT022338-045418 and GMRT021759-061642), 
325 MHz radio emission appears much extended than that at 1.4 GHz, 
suggesting the presence of diffuse relic emission detected only at the low-frequency.     
We note that the majority (33/48) of our remnant sources appear unresolved at both frequencies. 
In Figure~\ref{fig:RadioContUR} we show three examples of remnant candidates exhibiting unresolved emission 
$-$ (i) a bright unresolved source GMRT022211-054906 with S$_{\rm 325~MHz}$ = 121.4$\pm$0.3 mJy, 
(ii) an unresolved source GMRT021836-035711 with intermediate level of flux density S$_{\rm 325~MHz}$ = 17.4$\pm$0.2 mJy, 
and (iii) a faint unresolved source GMRT021646-051004 with S$_{\rm 325~MHz}$ = 1.93$\pm$0.12 mJy. 
We find that both resolved and unresolved sources are present at the wide range of flux densities with 
S$_{\rm 325~MHz}$ spanning from 1.93 mJy to 132.8 mJy with a median value of 11.2 mJy. 
\par
Also, we inspected 3 GHz Very Large Array Sky Survey (VLASS) quick look image cut-outs for 
all our remnant candidates. The VLASS provides wide-band (2 $-$ 4 GHz) images with median noise-rms of 0.145 mJy 
and angular resolution of 2$^{\prime\prime}$.5 \citep{Gordon21}.      
We find that only 10/48 of our remnant candidates show detection in the VLASS. 
One of our sample source GMRT020750-030455 exhibits a clear double lobe morphology with a total end-to-end projected 
size of 13$^{\prime\prime}$.6 ($\sim$ 112 kpc at $z$ = 1.04) in the VLASS image. 
Table~\ref{tab:sample} lists 3.0 GHz total flux densities derived from the VLASS epoch 1 quick look catalogue 
and the spectral index measured between 1.4 GHz and 3.0 GHz. 
We used only catalogue source components with duplicate$\_$flag $<$ 2 and quality$\_$flag = 0, and corrected for 10$\%$ 
systematic underestimation of flux density. 
The non-detection of a large fraction (38/48 = 79$\%$) of our remnant candidates, in particular for relatively 
bright sources at low-frequency (S$_{\rm 150~MHz}$ $>$ 20 mJy), in the VLASS can be understood 
if they posses relic emission of low-surface-brightness characterised with steep spectral index. 
The non-detection of relatively fainter remnant candidates can also be attributed to the 
shallow sensitivity of the VLASS. 
We note that, as expected, our remnant candidates continue to exhibit steep spectral index at higher frequency 
(1.4~GHz $-$ 3.0~GHz) regime. While, 05/10 of our remnant candidates detected in the VLASS show flatter 
spectral index (${\alpha}$ $>$ -0.5) in the higher frequency regime, despite exhibiting steeper spectrum at 
low-frequency regime ($<$ 1.4 GHz). The changing spectral characteristics of these remnant candidates can be 
understood if these candidate sources depict recurrent AGN activity such that a new phase of AGN activity 
showing flat or inverted radio spectrum at higher frequencies co-exists along with the relic emission from the previous 
episode of AGN activity showing steep spectrum at lower frequencies \citep[see][]{Murgia11}. 
Therefore, we note that a fraction of our remnant candidates can possibly show the existence of active core, if deep 
high-frequency observations are performed. 
\subsection{Redshifts}
\label{sec:redshift}
\begin{figure*}
\includegraphics[angle=0,width=8.0cm,trim={0.0cm 0.0cm 0.0cm 0.0cm},clip]{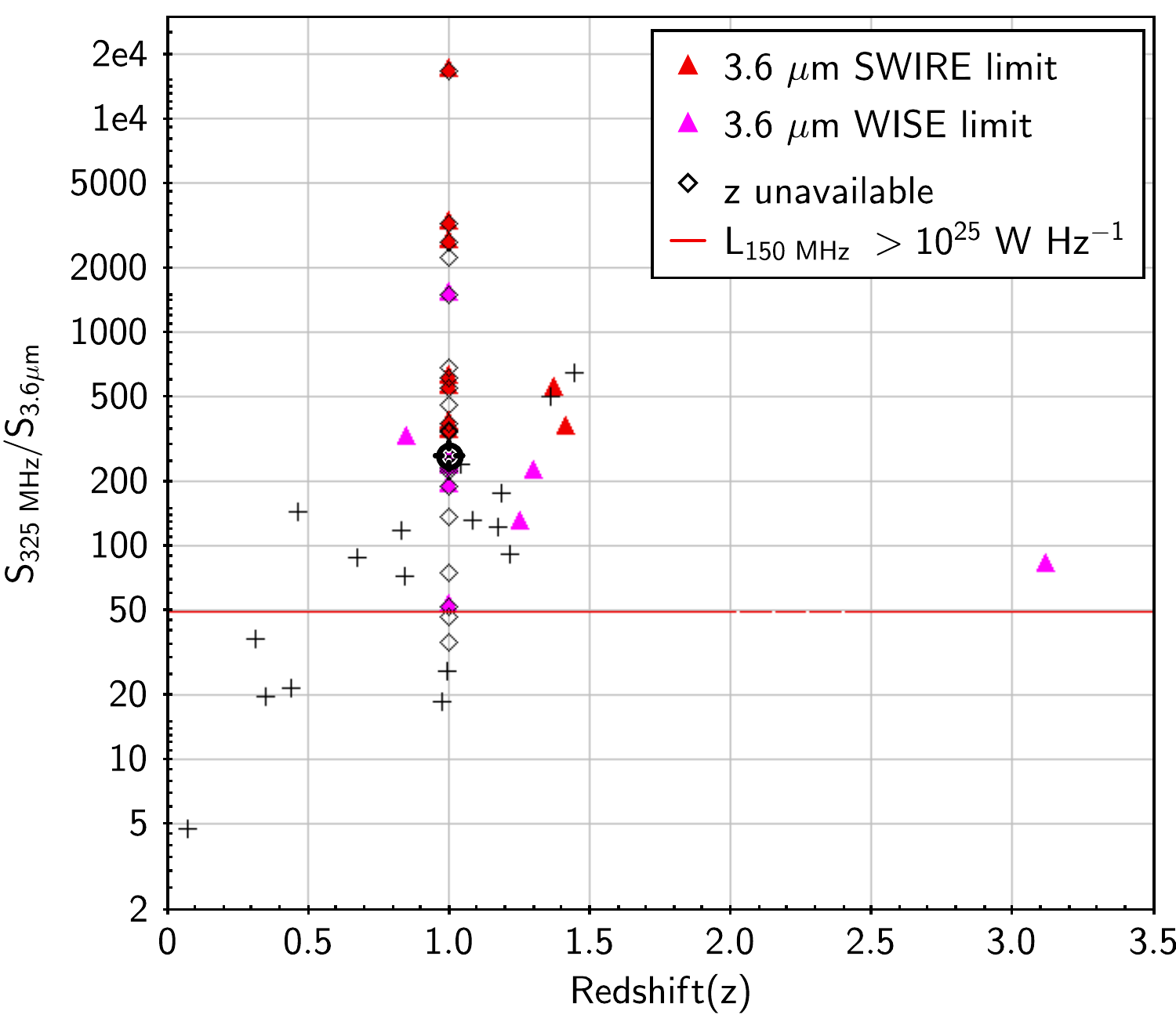}
\includegraphics[angle=0,width=8.0cm,trim={0.0cm 0.0cm 0.0cm 0.0cm},clip]{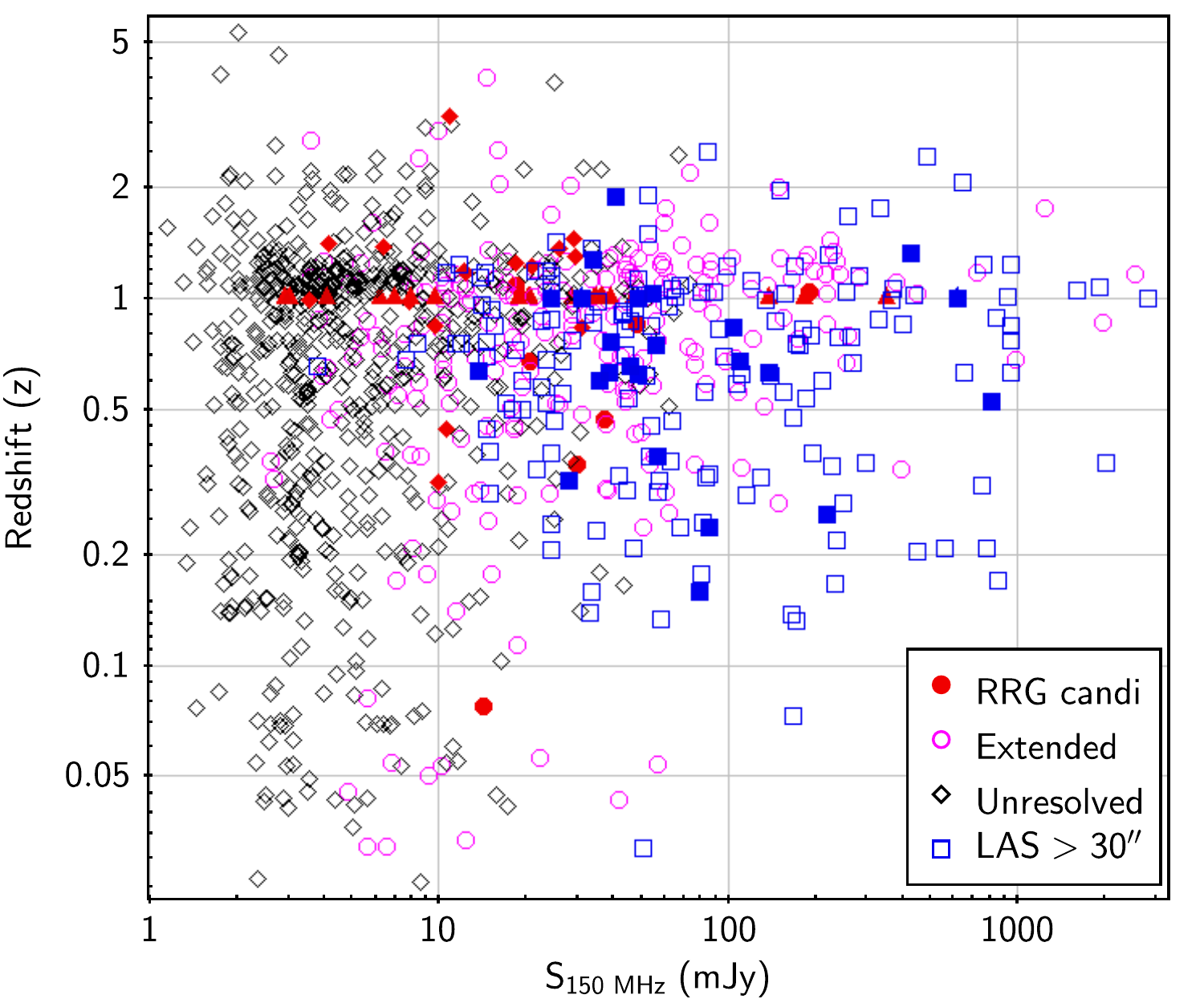}
\caption{{\it Left panel}: Ratio of 325 MHz radio flux density to 3.6~$\mu$m flux density plotted against the redshift. 
Sources with no redshift estimates are assigned an upper limit of $z$ $>$ 1.0 based the ratio of 325 MHz to 3.6$\mu$m flux densities 
similar to the high-$z$ radio sources. {\it Right panel} : 150 MHz flux density versus redshift plot. }
\label{fig:zplot} 
\end{figure*}

We find that only 858/1516 (56.6$\%$) of our sample sources have optical counterparts with redshift estimates in the 
HSC-SSP catalogues. 
Spectroscopic redshifts are available for 353 sources and remaining 505 sources have photometric redshifts.   
Among the 48 remnant candidates only 23 sources have redshift estimates that include spectroscopic redshifts for four sources and 
photometric redshifts for remaining 19 sources. 
We find that 23 remnant candidates have redshifts in the range of 0.077 to 3.12 with a median value of 1.04. 
While, 835 active radio sources have a wide redshift distribution ranging from 0.026 to 5.29 with a median value of 0.81. 
Thus, in compared to active sources, our remnant candidates are relatively at higher redshifts. 
The two-sample KS test shows that the redshift distributions of our remnant candidates and active sources are somewhat 
different with p-value = 0.13 (see Table~\ref{tab:stat}). 
Our 25 remnant candidates with no available redshift estimates have either undetected or too faint optical counterparts. 
We attempt to place a lower limit on the redshifts of these remnant candidates by using a diagnostic plot based 
on the ratio of 325~MHz flux density to 3.6~$\mu$m flux density (S$_{\rm 325~MHz}$/S$_{3.6~{\mu}m}$). 
In Figure~\ref{fig:zplot} we plot S$_{\rm 325~MHz}$/S$_{3.6~{\mu}m}$ 
versus redshift plot for our remnant candidates. 
We obtain 3.6~${\mu}$m flux density from the Spitzer Wide-area Infrared Extragalactic Survey 
(SWIRE; \cite{Lonsdale03}), whenever available. For sources falling outside the SWIRE coverage we check their 
detection in 3.4~$\mu$m band of the WISE all-sky imaging survey \citep{Wright10}. 
We convert 3.4~${\mu}$m flux density to 3.6~$\mu$m by assuming a power law spectrum with index of -2 in 
the near-IR regime. For sources with no detection in the SWIRE and WISE images we placed an upper limit 
on 3.6~${\mu}$m flux density based on the 5$\sigma$ detection limits. 
\par
From Figure~\ref{fig:zplot} it is evident that the ratio S$_{\rm 325~MHz}$/S$_{3.6~{\mu}m}$ increases with the 
increase in redshift, which is analogous to the well known correlation between K band magnitude and redshift 
(K$-z$ relation) for radio galaxies \citep{Willott03}. We note that all but one of our remnant candidates 
at $z$ $>$ 1.0 have ratio S$_{\rm 325~MHz}$/S$_{3.6~{\mu}m}$ $>$ 100. 
Hence, high ratio of S$_{\rm 325~MHz}$/S$_{3.6~{\mu}m}$ can be considered as 
an indicator for high redshift. Our remnant candidates with no detected 
3.6~$\mu$m counterparts have high upper limits on S$_{\rm 1.4~GHz}$/S$_{3.6~{\mu}m}$ in the range of 52 to 16632 with 
a median value 338, similar to H$z$RGs. In fact, six of the 20 remnant candidates with no detected 
3.6~$\mu$m counterparts have photometric redshift estimates in the range of 0.85 to 3.11 with a median value of 1.33. 
Hence, considering the fact that all of our remnant candidates with no redshift estimates show high values or limits of 
S$_{\rm 325~MHz}$/S$_{3.6~{\mu}m}$, similar to H$z$RGs or radio-loud AGN at $z$ $>$ 2 \cite{Singh17}, 
we place an upper limit of $z$ $>$ 1.0 for all our remnant candidates with no available redshifts.   
\begin{table*} 
\begin{minipage}{160mm}
\caption{Comparison between remnant candidates and active sources}
\label{tab:stat}
\begin{tabular}{ccccccccc}
\hline
Parameter  & \multicolumn{3}{c}{Remnant candidates}  &  \multicolumn{3}{c}{Active sources}  &  \multicolumn{2}{c}{KS test}   \\  \cline{2-4} \cline{5-7} \cline{8-9}
     &  N$_{\rm source}$     & range             &  median  &  N$_{\rm source}$   & range     &  median  &  D   & p-value  \\ \hline
S$_{\rm 150~MHz}$ (mJy)   & 48  & 2.9 $-$ 350.3  & 19.5   & 1468  & 1.15 $-$ 2543.3 & 5.8 & 0.32 &  2 $\times$ 10$^{-4}$  \\         
S$_{\rm 325~MHz}$ (mJy)                & 48  & 1.93 $-$ 132.8  & 11.2  & 1468  & 0.78 $-$ 1511.2 & 4.8 & 0.21  & 0.03  \\  
${\alpha}_{\rm 150~MHz}^{\rm 1.4~GHz}$ & 48  & -1.94 $-$ -0.86 & -1.04  & 1118 & -1.91 $-$ 1.09  &-0.61 & 0.81 & 1.6$ \times$ 10$^{-6}$ \\   
${\Delta}{\alpha}$                     & 48  & 0.50 $-$ 1.10   & 0.62  & 1118  & -3.81 $-$ 2.98  & 0.38  & 0.58 & 2.1$ \times$ 10$^{-4}$ \\
   $z$                                 & 23  & 0.077 $-$ 3.12  & 1.04  &  835  & 0.026 $-$ 5.29  & 0.81  & 0.25 & 0.13 \\  
LAS (kpc)                              & 7   &  27.1 $-$ 201.2 & 124   & 266 & 11.5 $-$ 245.1 & 157.4 & 0.52 & 0.032 \\  
logL$_{\rm 150~MHz}$ (W~Hz$^{-1}$)     & 23  & 23.33 $-$ 27.06 & 25.99 & 835 & 21.56 $-$ 28.29 & 25.17 & 0.46 & 2 $\times$ 10$^{-4}$ \\  
\hline
\end{tabular}
\\
{\bf Notes} - The two sample Kolmogorov$-$Smirnov (KS) test is non-parametric statistical test that examines 
the hypothesis that two samples come from same distribution. D represents the maximum difference between the cumulative 
distributions of two samples and p-value is the probability that the null hypothesis, i.e., two samples comes from
same distribution, is true. 
\end{minipage}
\end{table*} 	
\subsection{Radio luminosity versus radio size (P-D plot)}
\label{sec:PDplot}
\begin{figure*}
\includegraphics[angle=0,width=8.0cm,trim={0.0cm 0.0cm 0.0cm 0.0cm},clip]{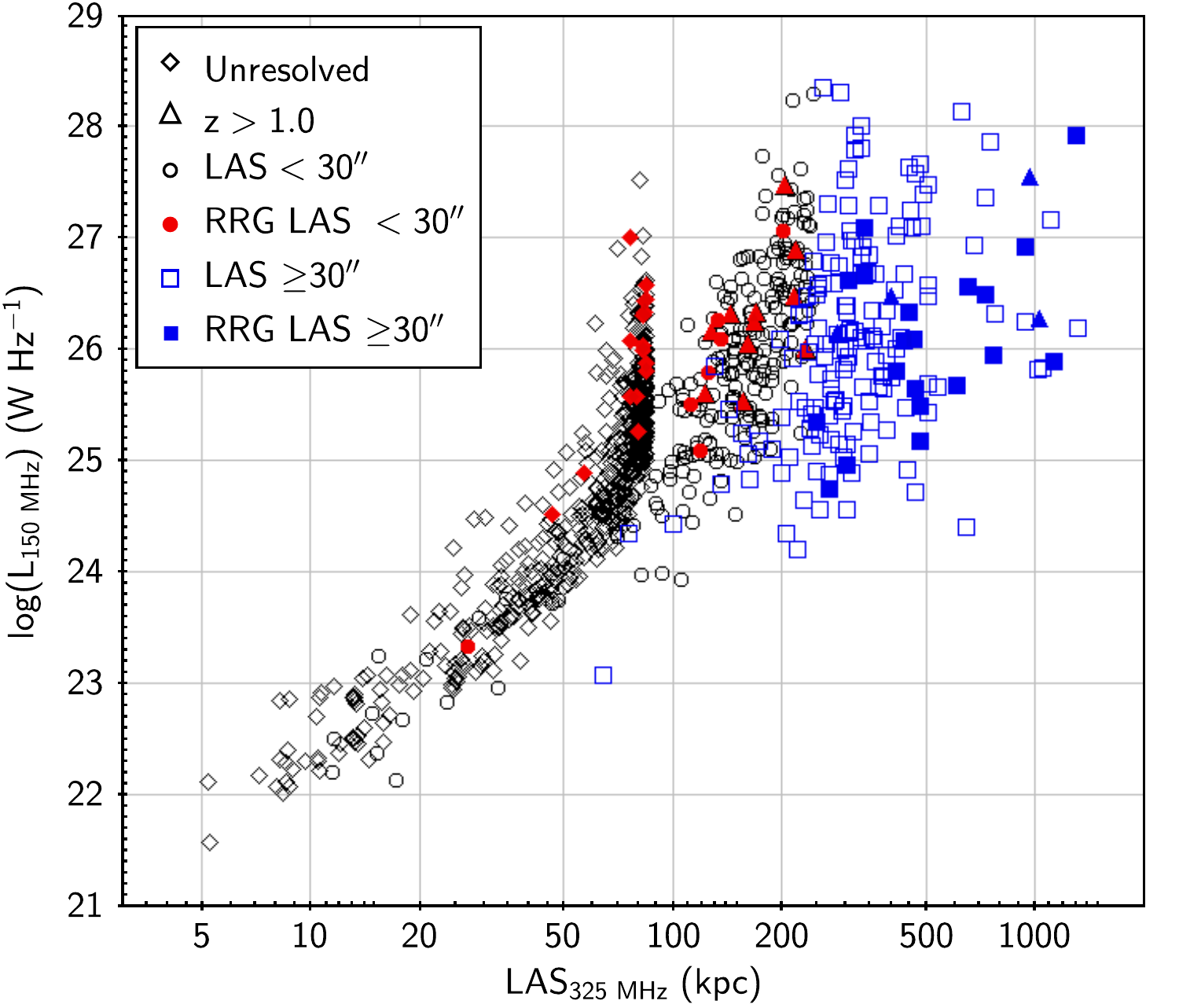}
\includegraphics[angle=0,width=8.0cm,trim={0.0cm 0.0cm 0.0cm 0.0cm},clip]{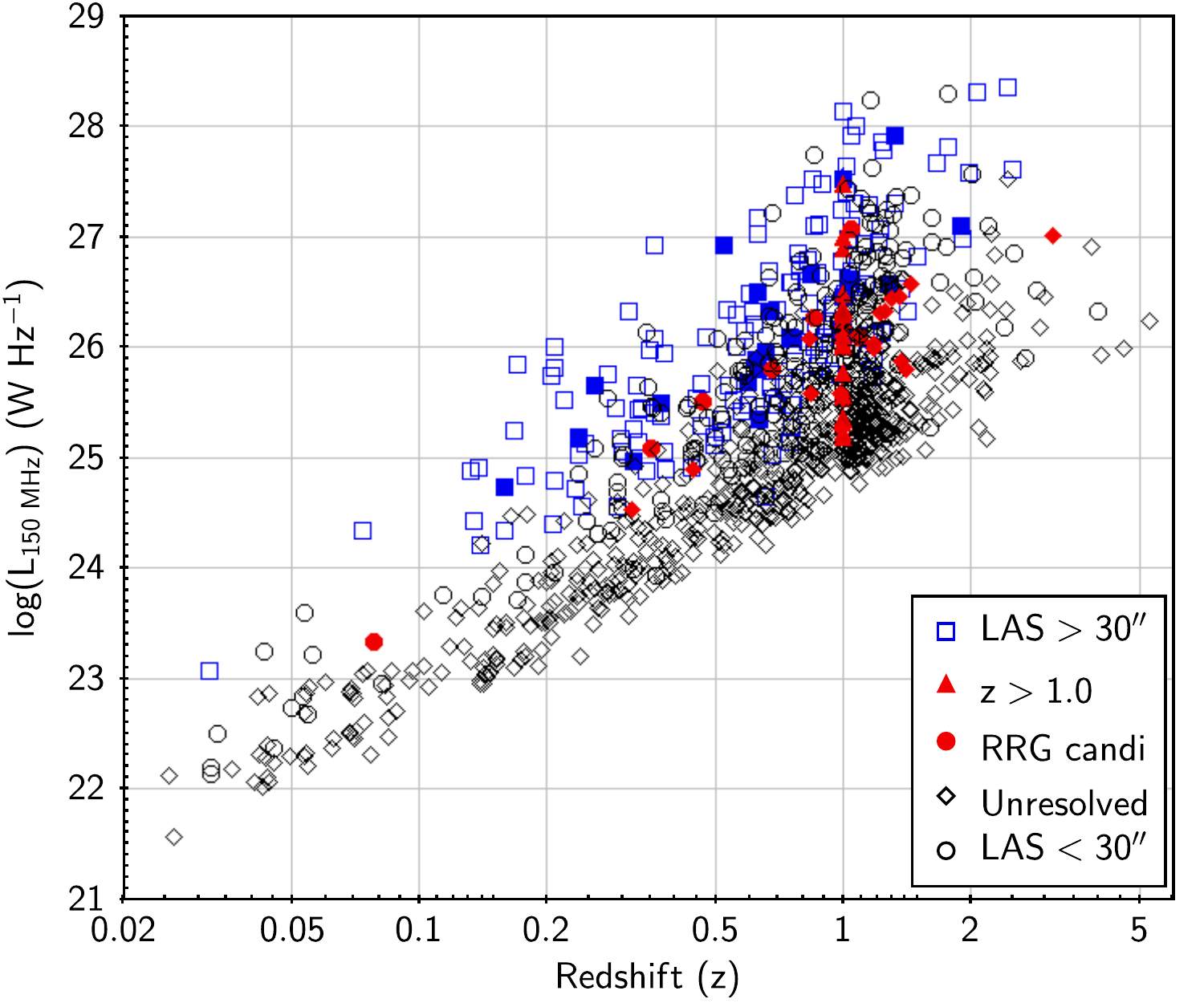}
\caption{{\it Left panel}: 150 MHz rest-frame radio luminosity versus radio size plot. For unresolved sources an upper limit of 10$^{\prime\prime}$ on the radio size is considered. {\it Right panel} : Redshift versus 150 MHz rest-frame radio luminosity.}
\label{fig:PDPlot} 
\end{figure*}

According to the dynamical evolutionary models radio size and radio luminosity of a radio galaxy continues 
to increase till it becomes fully evolved with total end-to end radio size in the range of 
a few hundreds of kpc to even a few Mpc \citep{An12}. 
Once AGN activity switches off radio luminosity begins to decrease due to radiative cooling. 
Therefore, in compared to active sources, RRGs are expected to show lower radio luminosities but larger radio sizes similar to that for 
fully evolved radio galaxies. We use radio luminosity versus radio size plot, commonly known as P$-$D plot, 
to infer the evolutionary stage of our remnant candidates and compare them with active sources. 
We derive 150 MHz luminosity for all our sources with available redshift estimates and 
apply K-correction using spectral index measured between 
150~MHz and 1.4~GHz (L$_{\rm 150~MHz}$ = 4$\pi$S$_{\rm 150~MHz}$ D$_{\rm L}^{2}$ (1+$z$)$^{1+{\alpha}}$, where D$_{\rm L}$ is 
luminosity distance and ${\alpha}$ = ${\alpha}_{\rm 150~MHz}^{\rm 1.4~GHz}$).
Figure~\ref{fig:PDPlot} ({\it Left panel}) shows the 150 MHz radio luminosity versus radio size plot for our sample sources.  
We note that the radio size estimates are available for only 395/1516 (26$\%$) of our sample sources that show emission 
on scales larger than the synthesized beam of 10$^{\prime\prime}$ in the 325 MHz GMRT survey. 
For remaining (1121/1516 = 74$\%$) unresolved sources we place an upper limit of 10$^{\prime\prime}$. 
It is worth to mention that 1.4 GHz JVLA and FIRST images of higher resolution (5$^{\prime\prime}$.0) can place 
a tighter constraint on the radio size of apparently unresolved sources, but it would result a bias 
towards sources of higher surface-brightness. Also, a substantial fraction (356/1516 $\sim$ 23$\%$; see Table~\ref{tab:sample}) of our 
sample sources remained undetected at 1.4 GHz. 
Therefore, we prefer to use radio size obtained with the 325 MHz GMRT survey.       
\par
We compare linear radio sizes of our remnant candidates and active sources by considering only 
sources that have angular size estimates and redshifts. Sources with angular size estimates but no redshift estimates have 
only upper limits on their linear radio sizes. Also, unresolved sources with no redshift estimates do not allow us to 
place any limit on their radio sizes, hence, these sources are discarded from our analysis.  
We note that the redshift estimates are available for only 226/395 sources that include seven remnant candidates. 
We find that seven remnant candidates have linear radio size in the range of 27.1 kpc to 201.2 kpc with a median value of 124 kpc. 
The  linear radio sizes of 219 active radio sources are distributed in the range of 11.5 kpc to 245.1 kpc with a median value of 155.4 kpc.  
Thus, we find that the linear radio sizes of our remnant candidates and active sources are similar. 
The two-sample KS test suggests that the distributions of 
radio sizes of active and remnant candidates not very different with D = 0.52 and p-value = 0.03 (see Table~\ref{tab:stat}), 
even when the sample size of remnant candidates is too small. 
Also, our remnant candidates with upper limit $z$ $>$ 1.0 corresponds to the upper limit of 200 kpc for the radio sizes 
(see Figure~\ref{fig:PDPlot}, {\it Left panel}). 
Thus, we find that our remnant candidates of small angular sizes are small ($<$ 200 kpc) in their physical sizes too. 
It is worth to point out that the remnant candidates of large sizes reported in D21 have radio sizes distributed in the range of 
241.6 kpc to 1301 kpc with a median value of 478.3 kpc (see blue points in Figure~\ref{fig:PDPlot}).  
Therefore, we conclude that the SPC criterion applied on the small angular size sources 
(LAS $<$ 30$^{\prime\prime}$) allows us to identify remnant candidates of small sizes ($\leq$200 kpc). 
\par
To understand the nature of small-size remnant candidates we examine their radio luminosities. 
We find that our 23/48 remnant candidates have 150 MHz radio luminosities in the range of 
2.14 $\times$ 10$^{23}$ W~Hz$^{-1}$ to 1.15 $\times$ 10$^{27}$ W~Hz$^{-1}$ with 
a median value of 6.17 $\times$ 10$^{25}$ W~Hz$^{-1}$. While, active sources have 150 MHz luminosities distributed in the 
range of 3.6 $\times$ 10$^{21}$ W~Hz$^{-1}$ to 1.94 $\times$ 10$^{28}$ W~Hz$^{-1}$ with a median value of 
1.48 $\times$ 10$^{25}$ W~Hz$^{-1}$. 
The two-sample KS test shows that the 150 MHz luminosity distributions of remnant candidates and active sources 
are different {\ie}probability that the two distributions come from same distribution is only 2 $\times$ 10$^{-4}$.  
We note that the active radio sources are on average less luminous and contain a substantial fraction 
of sources at lower luminosities down to 3.6 $\times$ 10$^{21}$ W~Hz$^{-1}$. Sources of low radio luminosities 
can be possibly be radio-quiet AGN and star-forming galaxies.    
We find that, unlike active radio sources, both extended as well as unresolved remnant candidates are luminous {\ie}all but 
one remnant candidates have L$_{\rm 150~MHz}$ $\geq$ 10$^{24}$ W~Hz$^{-1}$. 
Our 25/48 remnant candidates with upper limit on their redshifts ($z$) $>$ 1.0 have high 150 MHz luminosities 
(L$_{\rm 150~MHz}$ $>$ 10$^{25}$ W~Hz$^{-1}$).   
Also, we point out that 150 MHz radio luminosities of our remnant candidates are similar to those of large-size 
remnant candidates reported in D21. Thus, we find that our remnant candidates represent a population of 
small-size ($\leq$ 200 kpc) luminous radio sources. 
High 150 MHz radio luminosity of our remnant candidates can be understood if they are at high redshifts, and/or appear 
bright at low-frequency due to their steep radio spectra. 
In Section~\ref{sec:smallsize} we discuss the plausible reasons for their small size.  
\section{Remnant candidates of small sizes}
\label{sec:smallsize}
From P-D plot it is evident that, in general, our remnant candidates have small radio size ($<$ 200 kpc) but 
high radio luminosity (L$_{\rm 150~MHz}$ $\geq$ 10$^{25}$ W~Hz$^{-1}$). The trend shown by our remnant candidates in the 
P-D plot is rather unusual as remnants are expected to show relatively large radio size and low radio luminosity.    
We note that the time evolution of both luminosity and physical size depends on the jet kinetic power, 
longevity of active phase, and large-scale environment of a radio source \citep{Hardcastle19}. 
Assuming radio luminosity as the proxy for jet power, the progenitors of our remnant candidates 
are expected to possess high jet power.
Thus, small radio-size can be expected if our remnant candidates reside in the centers of cluster environments wherein 
dense surrounding medium confines the growth of radio source \citep{Murgia11}. 
We investigated the large-scale environments of our remnant candidates by checking their association with 
the clusters detected in the XMM-LSS region. We used optically-selected cluster catalogue derived from 
the HSC-SSP survey \citep{Oguri18} and X-ray selected cluster catalogue based on the deep {\em XMM-N} survey \citep{Adami18}. 
Deep multi-band images of the HSC-SSP survey detect clusters over a wide range of redshifts 0.1 $<$ $z$ $<$ 1.1 with 
photometric redshift accuracy of ${\Delta}$z/ (1+z) $<$ 0.01 and richness (N$_{\rm gal}$) $\geq$ 15 accounting 
galaxies brighter than 24 magnitude in $z-$band \citep{Oguri18}. 
The {\em XMM-N} survey detects hot X-ray emitting gas present in the Intra-Cluster Medium (ICM) 
with the flux limit of a few times of 10$^{-15}$ 
erg s$^{-1}$ cm$^{-2}$ in 0.5 $-$ 10 keV band within 1$^{\prime}.0$ aperture \citep{Adami18}. Follow-up optical 
spectroscopic observations found X-ray detected clusters to be distributed over 0.0 $<$ $z$ $<$ 1.2 with one cluster at $z$ $=$ 2.0. 
We find that only two of our remnant candidates GMRT021408-053456 and GMRT021528-044045 are 
associated with clusters at $z$ = 0.445 and $z$ = 0.348, respectively, within a matching radius of 1$^{\prime}$.0 and the redshift uncertainty of ${\Delta}$$z$/ (1+$z$) $<$ 0.01. Remnant candidates GMRT021408-053456 and GMRT021528-044045 lie at the projected distance of nearly 2$^{\prime\prime}$.2 (12.7 kpc) and 5$^{\prime\prime}$.4 (26.8 kpc) away 
from the optically assigned cluster centers, respectively. 
Since only 15 remnant candidates fall within the clusters redshift cutoff limit $z$ $<$ 1.2, the fraction of remnant 
candidates associated with clusters is merely 13$\%$ (02/15) in our sample. 
Our result is consistent with \cite{Jurlin21} who found that only a small fraction (23$\%$) of their remnants 
reside in the cluster environments. 
Hence, we conclude that our small size remnant candidates generally reside in a less dense environments.
\par
The small size of our remnant candidates showing high radio luminosity and residing in non-cluster environments 
can be plausible if the active phase of these radio sources is unusually short ($<$ 50 $\times$ 10$^{6}$ years). 
Also, we caution that the high monochromatic radio luminosity may not necessarily imply high jet kinetic power 
due to degeneracy caused by variable parameters such as magnetic field strength in lobes and spectral 
curvature \citep{Turner18b}.    
Thus, it is possible that progenitors of our remnant candidates can possess low power jets  
resulting into smaller radio size. 
Further, projection effect can also play a role {\ie}a remnant in which jet axis were closely aligned to the 
line-of-sight would appear smaller in size. 
In addition to above factors we cannot rule out that if some of our remnant candidates possess very 
low-surface-brightness emission that remained undetected in the 325 MHz GMRT images, and hence, 
we underestimate their radio sizes. 
\section{Remnant fraction}
\label{sec:fraction}
The fraction of remnant sources in a sample of radio galaxies can allow us to constrain the AGN duty cycle 
and remnant evolutionary models. We attempt to determine the remnant fraction ($f_{\rm rem}$) in our sample 
of small size (LAS $<$ 30$^{\prime\prime}$) radio sources. 
The small size of our sources does not allow us to characterise their radio morphologies in detail, and hence, we cannot rule out the 
possibility of some of them being active with the presence of a faint core. Also, the identification of remnants based on the 
spectral curvature can be erroneous due to the resolution bias {\ie}when 1.4 GHz flux density from the FIRST or JVLA observations 
of relatively higher resolution is used. 
Thus, remnant sources identified in our study are only candidates and allow us to place 
only an upper limit on $f_{\rm rem}$. With the identification of 48 remnant candidates in a sample of 1516 radio 
sources we obtain $f_{\rm rem}$ $\leq$ 3.2$\%$. The upper limit on the remnant fraction would reduce 
to 2.8$\%$ if we exclude five remnant candidates with possible recurrent AGN activity. 
We note that $f_{\rm rem}$ found in our sample of small-size sources is much lower than that reported 
for large-size radio sources.   
For instance, using absent-core criterion \cite{Mahatma18} found $f_{\rm rem}$ $\leq$ 9$\%$ in a 
sample of 127 bright (S$_{\rm 150~MHz}$ $\geq$ 80 mJy) and large (LAS $\geq$ 40$^{\prime\prime}$) radio sources detected in 
the 150 MHz LOFAR survey. 
In the Lockman Hole field \citep{Jurlin21} reported $f_{\rm rem}$ $\leq$ 8.0$\%$ in a sample of 158 sources with 
S$_{\rm 150~MHz}$ $\geq$ 40 mJy and size $\geq$ 40$^{\prime\prime}$. 
D21 obtained $f_{\rm rem}$ $\leq$ 9$\%$ in a sample of 268 sources with flux density limit cutoff of 6.0 mJy at 325 MHz 
that corresponds to 10 mJy at 150 MHz, and a size cutoff limit of 30$^{\prime\prime}$. 
With the availability of deep 1.4 GHz JVLA survey in 
the XMM-LSS-JVLA region D21 placed a tighter constraint on $f_{\rm rem}$ to be $\leq$ 5$\%$. 
A significantly lower value of $f_{\rm rem}$ $\leq$3.2$\%$ found in our sample consists of small-size sources 
($<$ 30$^{\prime\prime}$) can be attributed to various factors such as $-$ (i) a single criterion based on the spectral curvature fails to 
identify all remnants and selects only those remnants for which spectral break falls within the frequency coverage of 150 MHz to 1.4 GHz, 
(ii) remnants depicting the last evolutionary phase of radio galaxies are generally of larger size, and only a low fraction of remnants appear small as they fail to grow large, 
(iii) sources at fainter flux densities have substantial contamination from radio-quiet AGN and star-forming galaxies. 
In following subsection we discuss the effects of various biases. 
\subsection {Selection criteria bias}
Radio spectrum of a remnant source can be characterized with a broken or curved power law wherein break frequency (${\nu}_{\rm b}$) 
depends on the time elapsed since the cessation of AGN activity (see Section~\ref{sec:criteria}). 
As the relic plasma ages ${\nu}_{\rm b}$ shifts progressively towards the lower frequency. 
Our spectral curvature criterion used in the frequency range of 150 MHz to 1.4 GHz would miss 
old remnants with ${\nu}_{\rm b}$ $<$ 150 MHz as well as young remnants with ${\nu}_{\rm b}$ $>$ 1.4 GHz. 
Limitations of the spectral curvature criterion are evident from the fact that D21 identified only 09/24 
of their remnant candidates using spectral curvature criterion. 
Thus, spectral curvature criterion allows us to identify only a fraction 
of remnant candidates and not the full population. Hence, a low fraction of remnant (3.2$\%$) found in our study can partly be 
attributed to the use of only one criterion.    
\subsection{Flux density and luminosity bias}
Our 1516 sample sources have 325 MHz flux densities in the range of 0.78 mJy to 1511 mJy with a median value of 4.8 mJy, 
while 48 remnant candidates are distributed across 1.93 mJy to 132.0 mJy with a median of 11.2 mJy (see Table~\ref{tab:stat}). 
From Figure~\ref{fig:zplot} ({\it Right panel}) it is evident that our sample contains increasingly a large fraction of 
sources at the fainter end reaching down to 1.0 mJy. 
If we place a flux density limit of S$_{\rm 325~MHz}$ $\geq$ 6.0~mJy or S$_{\rm 150~MHz}$ $\geq$ 10~mJy, 
similar to that used by D21, we obtain only 34 remnant candidates among 655 radio sources, 
yielding $f_{\rm rem}$ to be $\leq$ 5.2$\%$. Thus,  one of the reasons for the low remnant candidates fraction 
($f_{\rm rem}$ $\leq$3.2$\%$) in our sample is flux density bias. 
This can be understood as the faint source population is likely to be dominated by non radio-loud AGN 
while remnant candidates belong to the radio-loud AGN population. The flux density bias also translates into luminosity bias. 
If we place a luminosity cutoff limit of L$_{\rm 150~MHz}$ $\geq$ 10$^{24}$ W~Hz$^{-1}$ we recover a high fraction 
22/23 (95.7$\%$) of our remnant candidates but a relatively lower fraction 680/835 (81$\%$) of active sources, 
where only sources with estimated radio luminosities or available redshifts are considered. 
The different recovery rates for remnant and active sources can be explained if remnant candidates are of higher luminosities 
and active sources contain a large fraction of low-luminosity sources from non radio-loud sources. 
Therefore, we demonstrate that a substantial contamination by non radio-loud population, in particular towards fainter flux densities, is one of the reasons for the low remnant fraction. 
\subsection{Redshift bias}
In our sample only 23/48 remnant candidates have redshift estimates distributed in the range of 0.077 to 3.12 with a median value of 1.04. 
For remaining 25/48 remnant candidates we place an upper limit of $z$ $>$ 1.0 based on the ratio of 325 MHz flux density 
to 3.6 $\mu$m flux density.   
Thus, we find that our remnant candidates tend to lie at higher redshift with 37/48 (77$\%$) sources at $z$ $>$ 1.0.
It is worth to point out that using the same data D21 found that their remnant candidates of large-size (LAS $\geq$ 30$^{\prime\prime}$) 
lie systematically at lower redshifts in the range of 0.139 to 1.895 with a median value of 0.65. 
Therefore, unlike large-size remnant candidates reported in D21 majority of our remnant candidates appear small due to higher 
redshifts. From Figure~\ref{fig:zplot} ({\it Right panel}) it is clear that 
many remnant candidates of small angular sizes identified in our study tend to lie at higher redshifts and fainter flux densities. 
Hence, our study probes a somewhat different phase space in the flux density versus redshift plot. 
Therefore, it is possible that the sample containing high$-z$ radio sources may have 
different remnant fraction.  
\subsection{Fraction in the XMM-LSS-JVLA region}
There are a total of 637 radio sources falling within the 5.0 deg$^{2}$ of the XMM-LSS-JVLA region, 
while only 25 remnant candidates are identified within this region. 
We obtain $f_{\rm rem}$ $<$ 3.9$\%$ in the XMM-LSS region which is little higher than that found for the full sample 
($f_{\rm rem}$ $<$ 3.2$\%$). Slightly higher fraction in the JVLA survey region can be due to the availability of 
deeper 1.4 GHz data (5$\sigma$ = 0.08 mJy). We note that all but four of 356 sources with no detected 1.4 GHz counterparts fall 
outside the XMM-LSS-JVLA region, and upper 
limits on 1.4 GHz flux density is based on the FIRST (5$\sigma$ = 1.0 mJy) and NVSS (5$\sigma$ = 2.5 mJy). 
It is fairly possible that several sources with no detected 1.4 GHz counterparts can 
have 1.4 GHz flux densities much lower than their upper limits, and hence, these sources can show strong spectral curvature $\geq$ 0.5, 
if 1.4 GHz flux density estimates are available. Therefore, a fraction of active sources having only 1.4 GHz flux density upper limits 
can possibly turn out to be remnant candidates. These sources remained unidentified in our study due to the unavailability of 1.4 GHz flux 
densities. Therefore, the remnant fraction ($f_{\rm rem}$ $<$ 3.9$\%$) 
obtained in the XMM-LSS-JVLA region provides a better constraint. If we place 6.0 mJy flux density 
cutoff limit, same as used by D21, we obtain only 13 remnant candidates yielding $f_{\rm rem}$ $<$ 5.4$\%$ (13/240)
that is similar to the one ($f_{\rm rem}$ $<$ 5$\%$) obtained by D21 for the extended sources 
(LAS $\geq$ 30$^{\prime\prime}$) in the XMM-LSS region.   
Therefore, we find similar remnant fractions for both small-size (LAS $<$ 30$^{\prime\prime}$) as well as for large-size 
(LAS $\geq$ 30$^{\prime\prime}$) sources once the bias introduced by flux density limit is taken into account.   
\section{Results and conclusions}
\label{sec:conclusions}
We carried out a search for RRG candidates of small angular sizes (LAS $<$ 30$^{\prime\prime}$) using deep 
multi-frequency radio surveys (150 MHz LOFAR, 325 MHz GMRT and 1.4 GHz JVLA, NVSS and FIRST) in the XMM-LSS field. 
Our study is the first attempt to perform a systematic search for RRGs of small sizes.  
Owing to the small angular sizes of our sample sources we exploit spectral curvature criterion and discover 48 remnant candidates 
exhibiting strong spectral curvature {\ie}${\alpha}_{\rm low}$ - ${\alpha}_{\rm high}$ $\geq$ 0.5; 
where ${\alpha}_{\rm low}$ = ${\alpha}_{\rm 150~MHz}^{\rm 325~MHz}$ 
and ${\alpha}_{\rm high}$ = ${\alpha}_{\rm 325~MHz}^{\rm 1.4~GHz}$. Main conclusions of our study are outlined below. 
\\
(i) Unlike most of the previous studies limited to large and bright remnant sources our study identified 
remnant candidates of small angular sizes (LAS $<$ 30$^{\prime\prime}$) that include faint sources with flux 
density reaching down to 1.0 mJy at 150 MHz. 
Thus, our study unveils remnant candidates at the faintest flux density regime than that reported earlier. 
\\
(ii) Using 1.4 GHz images of higher resolution mainly from the JVLA survey we find that a fraction 
(15/48 = 31$\%$) of our RRG candidates have extended double-lobe like radio morphology, while majority of RRG 
candidates appear unresolved or slightly resolved in the 325 MHz GMRT images of 10$^{\prime\prime}$ resolution. 
Our remnant candidates show steep radio spectral index with ${\alpha}_{\rm 150~MHz}^{\rm 1.4~GHz}$ distributed in 
the range of -1.94 to -0.86 with a median value of -1.04, which is similar 
to that found for large-size RRGs identified mainly with morphological criteria. 
\\
(iii) 3 GHz VLASS quick look image cutouts show the detection of only 10/48 remnant candidates. 
Some of our remnant candidates continue to show steep spectral index at higher frequency (1.4~GHz $-$ 3.0~GHz). 
While, five remnant candidates show a flatter spectral index at higher frequency regime inferring the possibility of 
recurrent AGN activity. Deep high-frequency imaging of our remnant candidates would be useful in deciphering their nature. Thus, our study presents a sample of remnant candidates for follow-up observations.
\\ 
(iv) Unlike large-size remnant candidates reported in the previous studies our remnant candidates are found at relatively higher redshifts. There are 23/48 remnant candidates with redshift estimates in the range of 0.077 to 3.12 
with a median value of 1.04, while, remaining 25 remnant candidates have upper limit of $z$ $>$ 1.0. 
\\ 
(v) Radio luminosity versus size plot shows that our remnant candidates are of smaller sizes ($\leq$ 200 kpc) 
but have high 150 MHz luminosities similar to that found for large-size remnant candidates. 
150 MHz luminosities of our 23 remnant candidates with redshift measurements are distributed 
in the range of 2.12 $\times$ 10$^{23}$ W~Hz$^{-1}$ to 1.15 $\times$ 10$^{27}$ W~Hz$^{-1}$ with a median value of 
9.77 $\times$ 10$^{25}$ W~Hz$^{-1}$. Remaining 25 remnant candidates with $z$ $>$ 1.0 have L$_{\rm 150~MHz}$ $>$ 10$^{25}$ W~Hz$^{-1}$. 
We find that only a small fraction (02/15 $\sim$ 13$\%$) of our remnant candidates reside in clusters, 
and hence, small radio size is unlikely to be caused by dense large-scale environment. 
We speculate that a relatively short active phase ($<$ 50 $\times$ 10$^{6}$ years) can plausibly limit 
the growth of radio source to the size of $<$ 200 kpc, although projection effect can also make apparent 
size smaller.    
\\
(vi) Our study allows us to place an upper limit on the remnant fraction ($f_{\rm rem}$) to be 3.2$\%$. 
In the XMM-LSS-JVLA region $f_{\rm rem}$ is slightly higher (3.9$\%$) due to the availability of deep 1.4 GHz JVLA survey. 
Therefore, it is fairly possible that a fraction of sources having only upper limits on 1.4 GHz flux density 
may turn out to be remnant candidates, if actual 1.4 GHz flux density is much lower than the upper limit. 
Further, we find that the a low value of $f_{\rm rem}$ can arise due to various factors such as the usage of single selection criterion, 
contamination from non radio-loud AGN at fainter flux densities, and redshifts bias. 
Notably, using a flux density cutoff of 6.0 mJy at 325 MHz we find $f_{\rm rem}$ $<$ 5.4$\%$ in the XMM-LSS-JVLA region, 
which is similar to that for large-size remnant candidates in the same region. 
Therefore, $f_{\rm rem}$ is nearly same for both small-size and large-size remnant candidates once bias introduced by the 
flux density limit is accounted for. 
%
%

\vspace{6pt} 




\authorcontributions{Conceptualization, methodology, software, validation, V.S., S.D., Y.W. and I.C.H.; formal analysis, V.S. and S.D.; investigation, V.S. and S.D.; resources, V.S., Y.W. and I.C.H.; data curation, V.S., S.D. Y.W., and I.C.H.; writing---original draft preparation,  V.S. and S.D.; writing---review and editing, V.S., S.D., Y.W. and I.C.H.; visualization, V.S., S.D., Y.W. and I.C.H.; supervision, V.S., Y.W. and I.C.H.; project administration, V.S., Y.W. and I.C.H. All authors have read and agreed to the published version of the manuscript.} 
\funding{This research received no external funding.}
%
%
\institutionalreview{Not applicable.}
%
%
\informedconsent{Not applicable.}
%
%
\dataavailability{The GMRT data reported in this study are available via the GMRT online archive https://naps.ncra.tifr.res.in/goa. 
Auxilary data from the LOFAR, JVLA, VLASS and HSC-SSP are available through the respective websites 
that are mentioned in the manuscript.} 
%
%
\acknowledgments{VS and SD acknowledge the support from Physical Research Laboratory, Ahmedabad, funded by the Department of Space, 
Government of India. CHI and YW acknowledge the support of the Department of Atomic Energy, Government of India, under project no.
12-R\&D-TFR5.02-0700. We thank the staff of GMRT who have made these observations possible. GMRT is run by the National
Centre for Radio Astrophysics of the Tata Institute of Fundamental Research. GMRT survey was carried out under program No. 4404-3
supported by the Indo-French Center for the Promotion of Advanced Research (Centre Franco-Indien pour la Promotion de la Recherche
Avancée).}
\conflictsofinterest{The authors declare no conflict of interest.} 






\end{paracol}
\reftitle{References}


\externalbibliography{yes}
\bibliography{RGCWpaper}

\end{document}